\documentclass[10pt,prc,amsmath,amssymb,twocolumn,aps,showpacs,superscriptaddress,groupedaddress]{revtex4-1}
\usepackage{graphicx}
\usepackage{color}

\begin{document}
%%%%%%%%%%%%%%%
\title{A realistic spectral function model for charged-current quasielastic-like neutrino and antineutrino cross sections on $^{12}$C}

%%%%%%%%%%%%%%%
\author{M.V. Ivanov}
\affiliation{Institute for Nuclear Research and Nuclear Energy, Bulgarian Academy of Sciences, Sofia 1784, Bulgaria}
\author{A.N.~Antonov}
\affiliation{Institute for Nuclear Research and Nuclear  Energy, Bulgarian Academy of Sciences, Sofia 1784, Bulgaria}
\author{G.D. Megias}
\affiliation{Departamento de F\'{i}sica At\'omica, Molecular y Nuclear, Universidad de Sevilla, 41080 Sevilla, Spain}
\author{J.A. Caballero}
\affiliation{Departamento de F\'{i}sica At\'omica, Molecular y Nuclear, Universidad de Sevilla, 41080 Sevilla, Spain}
\author{M.B.~Barbaro}
\affiliation{Dipartimento di Fisica, Universit\`{a} di Torino and INFN, Sezione di Torino, Via P. Giuria 1, 10125 Torino, Italy}
\author{J.E. Amaro}
\affiliation{Departamento de F\'{\i}sica At\'omica, Molecular y Nuclear, and Instituto de F\'{\i}sica Te\'orica y Computacional Carlos I, Universidad de Granada, Granada 18071, Spain}
\author{I. Ruiz Simo}
\affiliation{Departamento de F\'{\i}sica At\'omica, Molecular y Nuclear, and Instituto de F\'{\i}sica Te\'orica y Computacional Carlos I, Universidad de Granada, Granada 18071, Spain}
\author{T.W. Donnelly}
\affiliation{Center for Theoretical Physics, Laboratory for Nuclear Science and Department of Physics, Massachusetts Institute of Technology, Cambridge, Massachusetts 02139, USA}
\author{J.M. Ud\'{\i}as}
\affiliation{Grupo de F\'{\i}sica Nuclear, Departamento de Estructura de la Materia, F\'{\i}sica Aplicada y Electr\'{o}nica and UPARCOS, Universidad Complutense de Madrid, CEI Moncloa, 28040 Madrid, Spain}

%%%%%%%%%%%%%%
\begin{abstract}
 A detailed study of charged current quasielastic neutrino and antineutrino scattering cross sections on a $^{12}$C target with no pions in the final state is presented.   The initial nucleus is described by means of a realistic spectral function $S(p,{\cal E})$ in which nucleon-nucleon correlations are implemented by using natural orbitals through the Jastrow method. The roles played by these correlations and by final-state interactions are analyzed and discussed. The model also includes the contribution of weak two-body currents in the two-particle two-hole sector, evaluated within a fully relativistic Fermi gas.
  The theoretical predictions are compared with a large set of experimental data for double-differential, single-differential and total integrated cross sections measured by the MiniBooNE, MINER$\nu$A and T2K experiments. Good agreement with experimental data is found over the whole range of neutrino energies. The results are also in global good agreement with the predictions of the superscaling approach, which is based on the analysis of electron-nucleus scattering data, with only a few differences seen at specific kinematics.
\end{abstract}


\maketitle

\section{Introduction\label{sec:introduction}}

Having a good understanding of neutrino properties is presently one of the highest priorities in fundamental physics, explaining why considerable effort has been expended in recent years by a large number of researchers. Most of the recent (MiniBooNE, T2K, MINER$\nu$A, NOvA) and future (DUNE, HyperK) long baseline neutrino experiments make use of complex nuclear targets. Hence, precision measurement of neutrino oscillation parameters and the CP violation phase requires one to have excellent control over medium effects in neutrino-nucleus scattering. In fact, nuclear modeling has become the main issue in providing neutrino properties with high accuracy. A detailed report on the study of neutrino-nucleus cross sections is presented in the NuSTEC White Paper~\cite{Alvarez-Ruso:2017oui}.

In this work we restrict our attention to charged-current (CC) neutrino-nucleus scattering processes in the GeV region, and follow closely the analysis already presented in \cite{PhysRevD.94.093004} within the framework of the Superscaling approach. For a detailed discussion of the scaling and superscaling models, the reader is referred to~\cite{Amaro:2004bs, Amaro:2006tf,doi:10.1146/annurev.ns.40.120190.002041, PhysRevLett.45.871, PhysRevC.36.1208, PhysRevC.39.259, PhysRevC.43.1155, PhysRevC.46.1045, PhysRevC.53.1689, degliAtti1999447,PhysRevC.38.1801, Barbaro1998137, PhysRevLett.82.3212, PhysRevC.60.065502, PhysRevC.65.025502, PhysRevC.69.035502, PhysRevC.74.054603, PhysRevC.69.044321, PhysRevC.71.014317, PhysRevC.73.047302, PhysRevC.77.034612, RuizSimo:2018kdl}. The analysis of CC (anti)neutrino scattering with no pions in the final state (denoted as CC0$\pi$) has proven the essential role played by two-particle two-hole (2p-2h) Meson Exchange Currents (MEC) in addition to the quasielastic (QE) response. The inclusion of these two contributions has allowed one to explain data for different experiments without the need to modify the standard value of the axial mass $M_A$~\cite{Martini:2009uj, Amaro:2010sd, Nieves:2011yp, PhysRevC.86.014614, PhysRevD.94.093004}. It is important to point out that, contrary to electron scattering, in (anti)neutrino-nucleus processes the neutrino energy is not known precisely, and this implies that one- and two-body responses cannot be disentangled in the inclusive experimental data, where only the outgoing lepton is detected.

This paper complements the analysis already presented in~\cite{PhysRevD.94.093004}, but here the QE regime is described making use of realistic spectral functions instead of the superscaling prescription denoted as SuSAv2~\cite{PhysRevC.90.035501}. The spectral functions considered here account for effects linked to energy dependences and short-range nucleon-nucleon (\emph{NN}) correlations computed through the Jastrow method (see~\cite{PhysRevC.89.014607, PhysRevC.91.034607, PhysRevC.83.045504, PhysRevC.83.045504} for details). With regard to FSI, these are included by introducing a time-independent optical potential that describes the interaction between the struck nucleon and the residual nucleus. Concerning the treatment of 2p-2h excitations, we follow our previous studies in~\cite{PhysRevD.90.033012, Simo:2016ikv, PhysRevC.94.054610} that present a microscopic calculation by including a fully relativistic model for the weak charged-current MEC in both longitudinal and transverse channels, and with vector and axial-vector contributions. The present model is applied to CC0$\pi$ (anti)neutrino scattering processes on carbon measured by the MiniBooNE~\cite{miniboone, miniboone-ant}, NOMAD~\cite{Lyubushkin:2009}, T2K~\cite{PhysRevD.93.112012} and MINER$\nu$A~\cite{PhysRevLett.111.022502, PhysRevLett.111.022501, cherylthesis, PhysRevD.97.052002} experiments spanning an energy range from hundreds of MeV up to $100$~GeV. In all figures in the paper, as reference, the results of SuSAv2-MEC model~\cite{PhysRevD.94.093004} are also presented.

The theoretical scheme of the work is given in Sect.~\ref{sec:formalism}, which contains a brief description of the methods to obtain a realistic spectral function, the main relationships concerning CCQE (anti)neutrino-nucleus reaction cross sections and a short summary on the inclusion of 2p-2h ingredients. The results of the calculations and discussion are presented in Sect.~\ref{sec:analysis}. A summary of the work and our conclusions are given in Sect.~\ref{sec:conclusions}.

\section{General Formalism\label{sec:formalism}}

\subsection{Expression for the cross sections\label{subsec21}}

We consider the process where an incident beam of (anti)neutrinos with 4-momentum $K^{\mu }=(\epsilon ,\mathbf{k})$ scatters off a nuclear target and a charged lepton with 4-momentum $K^{\prime \mu }=(\epsilon ^{\prime},\mathbf{k}^{\prime })$ emerges. The 4-momentum transfer $Q^{\mu }=(\omega ,\mathbf{q})\equiv(\epsilon -\epsilon ^{\prime } ,\mathbf{k}-\mathbf{k}^{\prime })$
is spacelike: $-Q^{2}=q^{2}-\omega ^{2}>0$.

The CC (anti)neutrino-nucleus inclusive cross section in the target laboratory frame can be written in the form (see \cite{Amaro:2004bs, doi:10.1146/annurev.ns.40.120190.002041} for details)
\begin{equation}
\left [ \frac{d^{2}\sigma}{d\Omega dk^{\prime}}\right]_{\chi}= \sigma_{0}{\cal F}_{\chi}^{2},\label{c.c.cr.s.}
\end{equation}
where $\chi=+$ for neutrino-induced reactions (in the QE case, $\nu_{\ell}+n\rightarrow \ell^{-}+p$, where $\ell=e, \mu, \tau$) and $\chi=-$ for antineutrino-induced reactions (in the QE case, $\overline{\nu}_{\ell}+p\rightarrow \ell^{+}+n$). In Eq.~(\ref{c.c.cr.s.})
\begin{equation}
\sigma_0=
\frac{G_F^2\cos^2\theta_c}{2\pi^2}
\left(k^\prime \cos\frac{\tilde\theta}{2}\right)^2
\end{equation}
depends on the Fermi constant $G_F =1.16639 \times 10^{-5}$~GeV$^{-2}$, the Cabibbo angle $\theta_c$ ($\cos\theta_c=0.9741$), the outgoing lepton momentum $k^\prime$, and the generalized scattering angle $\tilde\theta$
\begin{equation}
\tan^2  \frac{\tilde\theta}{2} = \frac{|Q^2|}{4\epsilon\epsilon^\prime-|Q^2|} \,.
\end{equation}

The function
\begin{multline}
{\cal F}_{\chi}^{2}=[\widehat{V}_\text{CC}R_\text{CC}+2\widehat{V}_\text{CL}R_\text{CL}+\widehat{V}_\text{LL}R_\text{LL}+\widehat{V}_\text{T}R_\text{T}]\\
+\chi[2\widehat{V}_\text{T$'$}R_\text{T$'$}] \label{new33}
\end{multline}
in Eq.~(\ref{c.c.cr.s.}) depends on the nuclear structure and is presented as a generalized Rosenbluth decomposition~\cite{Amaro:2004bs} containing leptonic kinematical factors, $V_K(q,\omega,\tilde\theta)$, and five nuclear response functions, $R_K(q,\omega)$, namely $VV$ and $AA$ charge-charge ($CC$), charge-longitudinal ($CL$), longitudinal-longitudinal ($LL$) and transverse ($T$) contributions, and $VA$ transverse ($T'$) contributions, where $V$($A$) denotes vector(axial-vector) current matrix elements. These are specific components of the nuclear tensor $W^{\mu\nu}$ in the QE region and can be expressed in terms of the superscaling function $f(\psi)$ (see~\cite{Amaro:2004bs} for explicit expressions).

\subsection{Models: HO+FSI, NO+FSI, and SuSAv2\label{subsec22}}

We consider three different theoretical calculations. Two of them, denoted as HO (harmonic oscillator) and NO (natural orbitals), make use of a spectral function $S(p,{\cal E})$, $p$ being the momentum of the bound nucleon and ${\cal E}$ the excitation energy of the residual nucleus, coinciding with the missing energy $E_m$ up to a constant offset~\cite{Barbaro1998137}. Both models include final-state interactions (FSI). The third model, SuSAv2, is instead based on the relativistic mean field (RMF) model and accounts consistently for both initial- and final-state interactions.

For the two models based on the spectral function (SF) we adopt the following procedure:
\begin{itemize}

\item[(i)] The spectral function $S(p,{\cal E})$ is constructed in the form~\cite{PhysRevC.83.045504, PhysRevC.89.014607, PhysRevC.91.034607}:
 \begin{equation}\label{HF+lorent}
    S(p,{\cal E})=\sum_{i}2(2j_i+1)n_i(p) L_{\Gamma_i}({\cal E} - {\cal E}_i),
\end{equation}
where the Lorentzian function is used:
\begin{equation}\label{lorent}
L_{\Gamma_i}({\cal E}-{\cal E}_i) = \dfrac{1}{\pi}\dfrac{\Gamma_i/2}{({\cal E}-{\cal E}_i)^2+(\Gamma_i/2)^2}\,
\end{equation}
${\Gamma_i}$ being the width of a given hole state. In Eq.~(\ref{HF+lorent}) we assume that proton and neutron shells for the same quantum numbers have the same momentum distribution and energies.

\item [(ii)] In Eq.~(\ref{HF+lorent}) the single-particle (s.p.) momentum distributions $n_i(p)$ are taken to correspond to harmonic-oscillator shell-model s.p. wave functions (in the case of HO model), or to natural orbitals s.p. wave functions $\varphi_\alpha (\mathbf{r})$ (in the case of NO model). The latter are defined in~\cite{Lowdin:1955} as the complete orthonormal set of s.p. wave functions that diagonalize the one-body density matrix $\rho(\mathbf{r},\mathbf{r'})$:
    \begin{equation}
    \rho (\mathbf{r},\mathbf{r}^{\prime} )=\sum_{\alpha} N_{\alpha} \varphi_{\alpha}^{*}(\mathbf{r}) \varphi_{\alpha}(\mathbf{r}^{\prime}) ,\label{defNO}
    \end{equation}
    where the eigenvalues $N_{\alpha}$ ($0\leq N_{\alpha}\leq 1$, $ \sum_{\alpha} N_{\alpha}=A$) are the natural occupation numbers. We use $\rho(\mathbf{r},\mathbf{r'})$ obtained within the lowest-order approximation of the Jastrow correlation methods~\cite{Stoitsov:1993}.

  \item [(iii)] The Lorentzian function [Eq.~(\ref{lorent})] is used for the energy dependence of the spectral function with parameters $\Gamma_{1p} = 6$~MeV, $\Gamma_{1s} = 20$~MeV, which are fixed to the experimental widths of the $1p$ and $1s$ states in $^{12}$C nucleus~\cite{Dutta:1999}.  The corresponding spectral function $S(p,{\cal E})$
    is presented in Fig.~2 of ~\cite{PhysRevC.91.034607}, where the two shells $1p$ and $1s$ are clearly visible.

\item[(iv)] For  given momentum transfer  $q$ and energy of the initial electron $\epsilon$ we calculate the electron-nucleus ($^{12}$C) cross section by using the PWIA expression for the inclusive electron-nucleus scattering cross section
\begin{multline}\label{cr.s.}
\frac{d\sigma_t}{d\omega d |\mathbf{q}|}={2\pi\alpha^2}\frac{ |\mathbf{q}|}{\epsilon^2} \int d{E}\:d^3p\:\frac{S_t(\mathbf{p}, {E})}{E_{\mathbf{p}}E_{\mathbf{ {p'}}}}\\\times\delta\big(\omega+M-{E}-E_{\mathbf{p'}}\big)L_{\mu\nu}^\text{em}H^{\mu\nu}_{\text{em, }t}\,.
\end{multline}
In Eq.~(\ref{cr.s.}) the index $t$ denotes the nucleon isospin, $L_{\mu\nu}^{\text{em}}$ and $H^{\mu\nu}_{\text{em, }t}$ are the leptonic and hadronic tensors, respectively, and $S_t(\mathbf{p}, E)$ is the proton (neutron) spectral function. The terms $E_{\mathbf{p}}$, $E_{\mathbf{{p'}}}$, and ${E}$ represent the energy of the nucleon inside the nucleus, the ejected nucleon energy, and the removal energy, respectively (see ~\cite{PhysRevC.77.044311} for details).

\item[(v)] Following the approach of ~\cite{PhysRevC.22.1680,PhysRevC.77.044311}, we account for the FSI of the struck nucleon with the spectator system  by means of a time-independent optical potential (OP): $U=V-\imath W$. In this case the energy-conserving $\delta$-function in Eq.~(\ref{cr.s.}) is replaced by
    \begin{multline}\label{deltaf}
    \delta(\omega+M-E-E_{\mathbf{p'}}) \rightarrow\\
    \rightarrow \dfrac{W/\pi}{W^2+[\omega+M-E-E_{\mathbf{p'}}-V]^2} ,
    \end{multline}
    with $V$ and $W$ obtained from the Dirac OP~\cite{PhysRevC.73.024608}.

  \item[(vi)] The corresponding scaling function $F(q,\omega)$ is calculated
as
\begin{equation} F(q,\omega) =
\dfrac{\left[d\sigma/d\epsilon'd\Omega'\right]_{(e,e')}}{\overline{\sigma}^{eN}
(q,\omega;p=|y|,{\cal E}=0)}\,, \label{scaling}
\end{equation}
where the electron single-nucleon cross section $\overline{\sigma}^{eN}$ is taken at $p = |y|$, the scaling variable $y$ being the smallest possible value of $p$ in electron-nucleus scattering for the smallest possible value of the excitation energy (${\cal E} = 0$). By multiplying $F(q,\omega)$ by $k_F$ the superscaling function $f(\psi)$ is obtained, where the scaling variable  $\psi$ has been introduced (see \cite{PhysRevC.38.1801,Barbaro1998137}). Similarly to $y$, $\psi$ is related to the minimum kinetic energy a nucleon must have to participate in the scattering reaction.

In Fig.~5 of ~\cite{PhysRevC.83.045504} the evolution of the superscaling function $f(\psi)$ for different values of $q$ from $100$ to $2000$~MeV/c was presented. Results were obtained making use of the HO momentum distributions for the 1p and 1s shells in $^{16}$O. It can be seen that for $q > 600-700$~MeV/c, scaling of the first kind is fulfilled [\emph{i.e.}, for high-enough values of the momentum transfer the explicit dependence of  $f (\psi)$ on $q$ is very weak]. Similar results are obtained using HO and NO momentum distributions for the 1p and 1s shells in $^{12}$C.

\item[(vii)] We implement Pauli blocking (PB) effects in the scaling function using the procedure proposed in~\cite{ROSENFELDER1980188}, which was applied in the SuSA approach~\cite{PhysRevD.89.093002}. The prescription consists in subtracting from the scaling function $f(\psi(\omega,q))$ its mirror function $f(\psi(-\omega,q))$.

\item[(viii)] Finally, the nuclear responses appearing in Eq.~(\ref{new33}) are calculated by multiplying $f(\psi)$ by the appropriate single-nucleon functions given in \cite{Amaro:2004bs}.

    \end{itemize}

\begin{figure}[htb]
\centering\includegraphics[width=8.5cm]{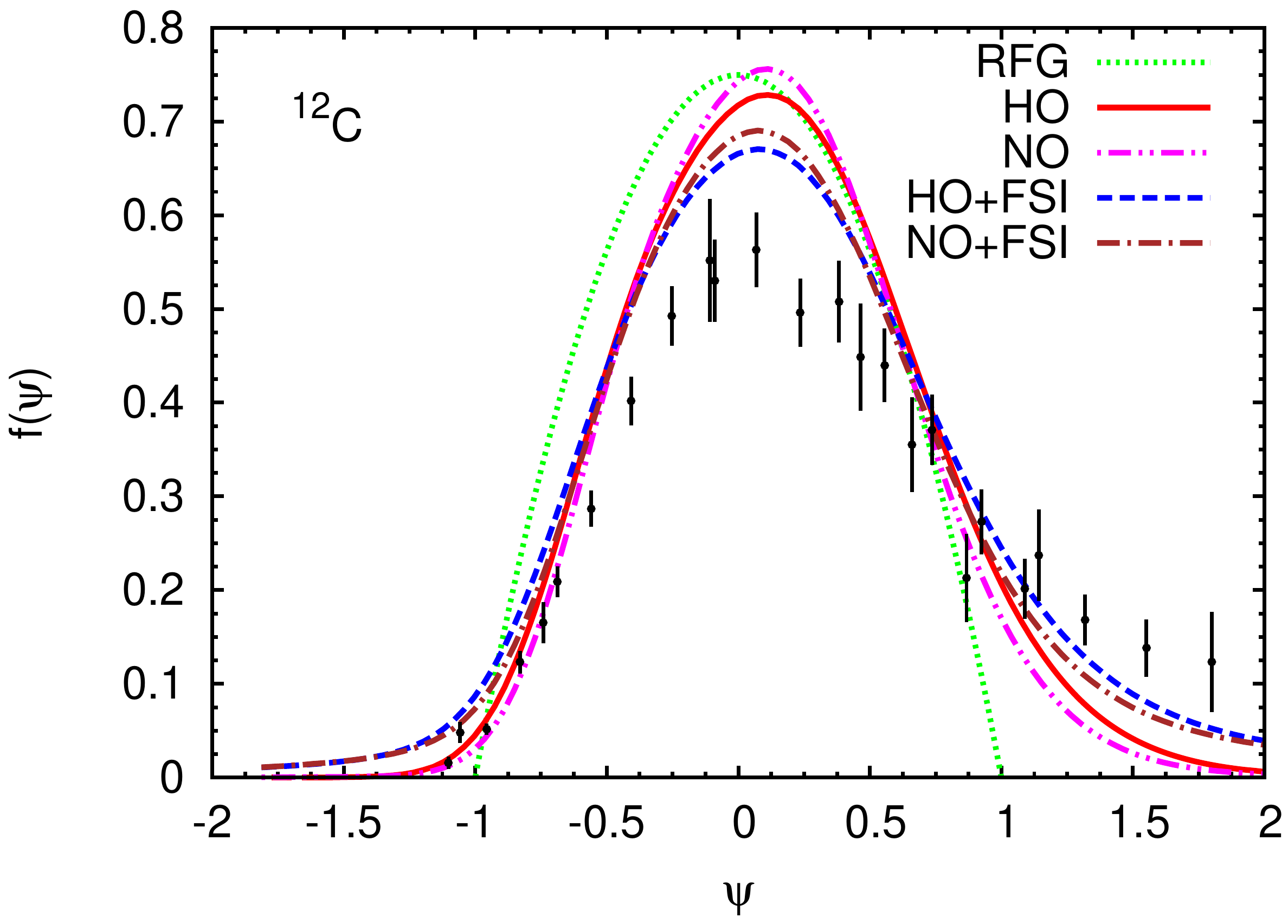}
\caption{(Color online) Results for the superscaling function $f(\psi)$ for $^{12}$C obtained using HO and NO approaches with (HO+FSI and NO+FSI) and without FSI (HO and NO) are compared with the RFG results, as well as with the longitudinal experimental data. \label{fpsi}}
\end{figure}

In Fig.~\ref{fpsi} the results for the superscaling function $f(\psi)$ within the HO+FSI and NO+FSI models are presented. As a reference are shown also the superscaling functions obtained without FSI and in the  RFG model, as well as the longitudinal experimental data~\cite{Amaro:2004bs}, where the 2p-2h contribution is zero or very small. Accounting for FSI leads to a redistribution of the strength, with lower values of the scaling function at the maximum and an asymmetric shape around the peak position, {\it viz.,} when $\psi=0$. Also, we see that the asymmetry in the superscaling function gets larger by using the Lorentzian function [Eq.~(\ref{lorent})] for the energy dependence of the spectral function than by using the Gaussian function~\cite{PhysRevC.83.045504, PhysRevC.89.014607, PhysRevC.91.034607}. The two spectral function models, including FSI, clearly give a much more realistic representation of the data than the relativistic Fermi gas. Few models, for example the relativistic mean field (RMF), are able to explain entirely the experimental data. In fact most models lie above the data similarly to the RFG and cannot explain the asymmetric shape. A recent calculation by J.E.~Sobczyk~\emph{et al.}~\cite{PhysRevC.97.035506}, based on a spectral function model, provides a scaling function which is very similar to ours, except for the low-momentum transfers.

Recently, an improved version of the superscaling prescription, called SuSAv2~\cite{PhysRevC.90.035501}, has been developed by incorporating RMF effects~\cite{PhysRevLett.95.252502, PhysRevC.74.015502, CABALLERO2007366} in the longitudinal and transverse nuclear responses, as well as in the isovector and isoscalar channels. This is of great interest in order to describe CC neutrino reactions that are purely isovector. Note that in this approach the enhancement of the transverse nuclear response emerges naturally from the RMF theory as a genuine relativistic effect.

The detailed description of the SuSAv2 model can be found in ~\cite{PhysRevD.94.013012, PhysRevC.90.035501, PhysRevD.94.093004}. Here we just mention that it has been validated against all existing $(e,e')$ data sets on $^{12}$C, yielding excellent agreement over the full range of kinematics spanned by experiments, except for the very low energy and momentum transfers, where all approaches based on impulse approximation (IA) are bound to fail. Furthermore, the success of the model depends on the inclusion of effects associated with two-body electroweak currents, which will be briefly discussed in the next Section.

\subsection{2p-2h MEC contributions\label{subsec23}}

Ingredients beyond the impulse approximation (IA), namely 2p-2h MEC effects, are essential in order to explain the neutrino-nucleus cross sections of interest for neutrino oscillation experiments~\cite{Martini:2009uj,PhysRevD.94.093004,Amaro:2010sd,Nieves:2011yp,PhysRevC.86.014614,Katori:2016yel,Alvarez-Ruso:2017oui}. In particular, 2p-2h MEC effects produce an important contribution in the ``dip'' region between the QE and $\Delta$ peaks, giving rise to a significant enhancement of the impulse approximation responses in the case of inclusive electron- and neutrino-nucleus scattering processes. In this work we make use of the 2p-2h MEC model developed in ~\cite{Simo:2016ikv}, which is an extension to the weak sector of the seminal papers \cite{VanOrden:1980tg, DePace:2003spn, Amaro:2010iu} for the electromagnetic case. The calculation is entirely based on the RFG model, and it incorporates the explicit evaluation of the five response function involved in inclusive neutrino scattering. The MEC model includes one-pion-exchange diagrams derived from the weak pion production model of ~\cite{Hernandez:2007qq}. This is at variance with the various scaling approaches that are largely based on electron scattering phenomenology, although also inspired in some cases by the RMF predictions.

Following previous works~\cite{PhysRevD.94.013012,PhysRevD.94.093004,Megias:2014qva,Ivanov:2015aya}, here we make use of a general parametrization of the MEC responses that significantly reduces the computational time. Its functional form for the cases of $^{12}$C and $^{16}$O is given in \cite{PhysRevD.94.013012, PhysRevD.94.093004, Megias:2017cuh}, and its validity has been clearly substantiated  by comparing its predictions with the complete relativistic calculation. The main merit of this procedure is that it can  easily be incorporated into the Monte Carlo neutrino event generators used in the analysis of neutrino oscillation experiments.

It should be noticed that the ground state used in the calculation of the 2p-2h is the RFG, while the spectral function ground state is used for the one-body response. The 2p-2h contribution for relativistic excitation energies is very involved. Already in the simplest model used here, the RFG, the calculation involves 7D numerical integrals and using the spectral function ground state is beyond the scopes of the present work. However, it has been shown~\cite{RUIZSIMO2017193} that the 2p-2h response functions of a Fermi gas are very similar to those of a bound system such as those obtained in the continuum shell model~\cite{AMARO1994365}, except for very low energy transfer (threshold for 2p-2h excitation). In the same reference it is also shown that for high-momentum transfer the 2p-2h responses are not sensitive to the fine details of the bound nucleon orbits. Furthermore, in these conditions the frozen nucleon approximation, where the two initial nucleons are considered at rest, can safely be used to compute the 2p-2h (provided that a smeared propagator for the delta excitation is used in the MEC operator), as shown in Ref.~\cite{RUIZSIMO2017193}. Hence the 2p-2h responses should not depend strongly on the spectral distribution of the bound nucleons.

We also note that the use of the RFG in the calculation of the 2p-2h response is common to other approaches~\cite{Martini:2009uj, NIEVES201390}.

\section{Analysis of results\label{sec:analysis}}

In this section we show the predictions of the two spectral function approaches previously described, HO and NO, both including FSI and 2p--2h MEC. We compare the results with data from different experiments: MiniBooNE, MINER$\nu$A, and T2K. Our study is restricted to the QE-like regime where the impulse approximation in addition to the effects linked to the 2p-2h meson-exchange currents play the major role. We follow closely the general analysis presented in \cite{PhysRevD.94.093004} for the case of the superscaling approach. Hence, for reference, we compare our new theoretical predictions with the results corresponding to the  SuSAv2-MEC model.

\begin{figure}[htb]
\centering
\includegraphics[width=8.5cm]{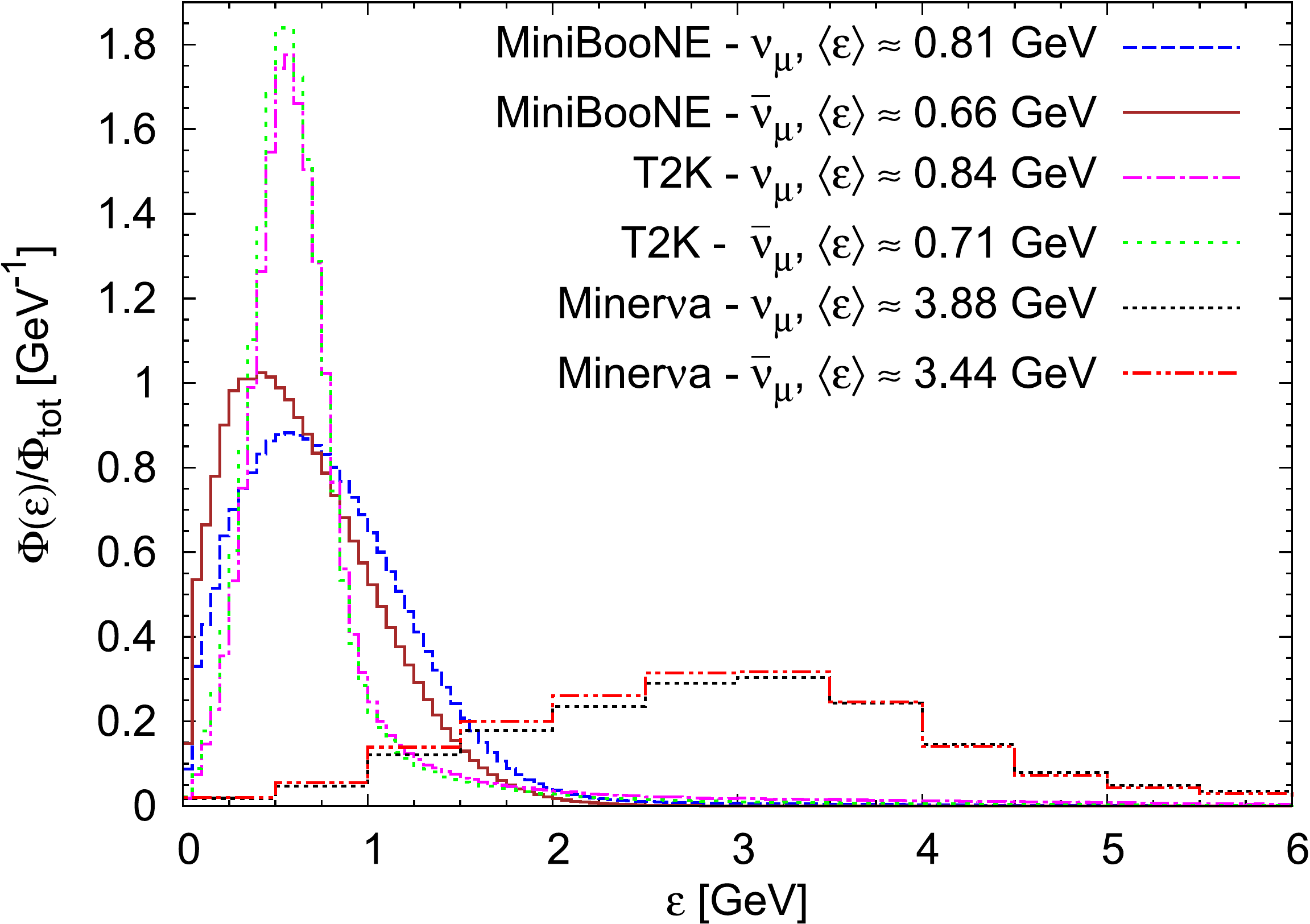}
\caption{(Color online) The predicted $\nu_\mu$ ($\overline{\nu}_\mu$) fluxes at the MiniBooNE~\cite{PhysRevD.79.072002}, T2K\cite{PhysRevD.87.012001}, and  MINER$\nu$A~\cite{PhysRevD.94.092005} detectors and corresponding mean energies.\label{fig:fluxes}}
\end{figure}

The predicted $\nu_\mu$ and $\overline{\nu}_\mu$ fluxes at the MiniBooNE~\cite{PhysRevD.79.072002}, T2K~\cite{PhysRevD.87.012001}, and MINER$\nu$A~\cite{PhysRevD.94.092005} detectors and corresponding mean energies are compared in Fig.~\ref{fig:fluxes}. $\Phi_\text{tot}$ is the total integrated $\nu_\mu$ ($\overline{\nu}_\mu$) flux factor:
\begin{equation}
\Phi_\text{tot}=\int\Phi(\epsilon)d\epsilon,
\end{equation}
where $\epsilon$ is incident beam energy. As observed, the neutrino and antineutrino mean energies corresponding to MiniBooNE and T2K experiments are rather similar, although the T2K energy flux shows a much narrower distribution. This explains the different role played by 2p-2h MEC effects in the two cases, these being larger for MiniBooNE (see \cite{PhysRevD.94.093004} and results in next sections). On the contrary, the MINER$\nu$A energy flux is much more extended to higher energies, with an average value close to $3.5-4.0$~GeV.

\subsection{MiniBooNE\label{subsec31}}

\begin{figure*}[htb]
\centering
\includegraphics[width=0.99\textwidth]{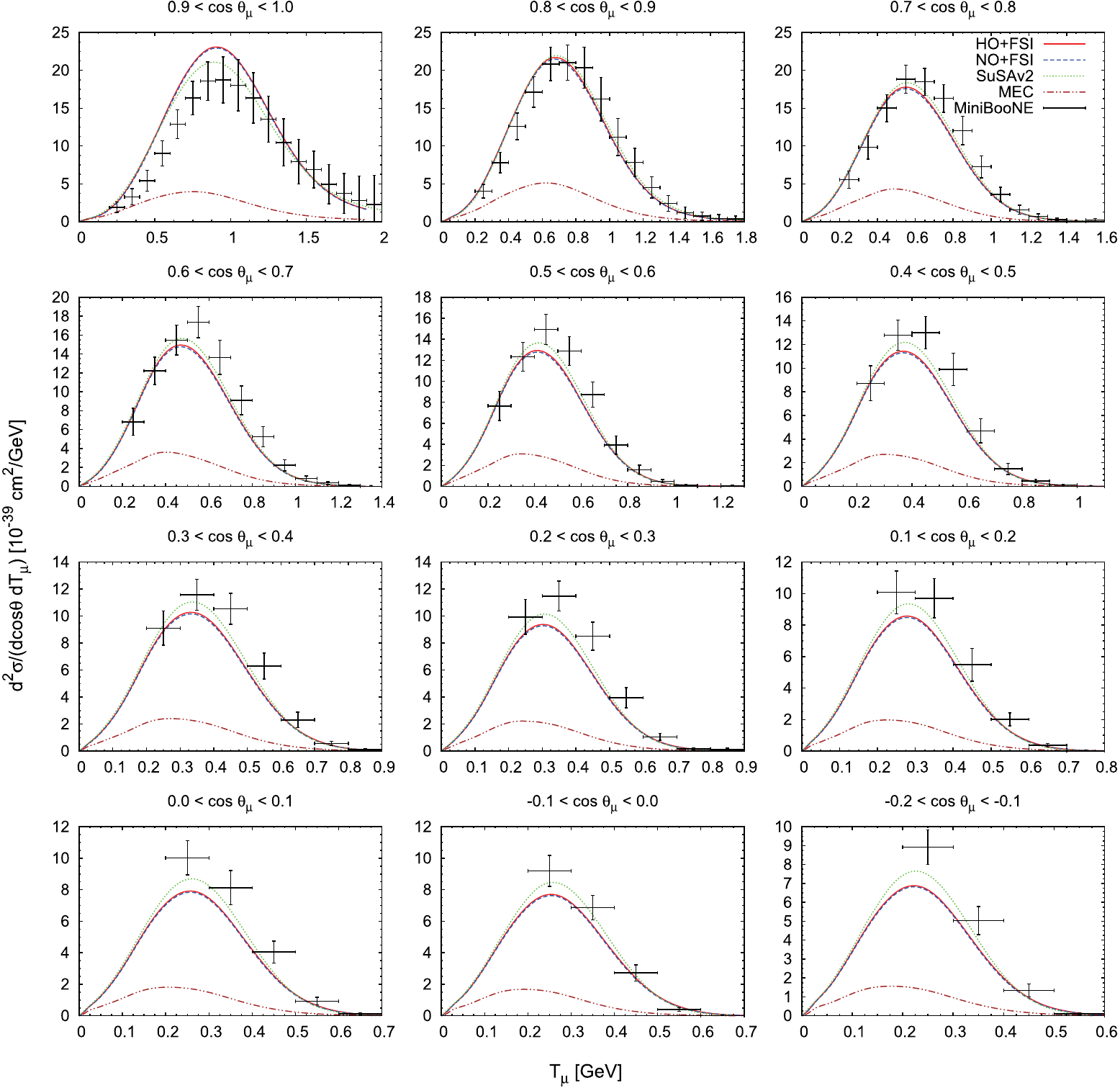}
\caption{(Color online) MiniBooNE flux-folded double differential cross section per target neutron for the $\nu_\mu$ CCQE process on $^{12}$C displayed versus the $\mu^{-}$ kinetic energy $T_\mu$ for various bins of $\cos \theta_\mu$ obtained within the SuSAv2, HO+FSI, and NO+FSI approaches including MEC. 2p--2h MEC results are shown separately. The data are from~\cite{miniboone}.\label{nuMiniBooNE1}}
\end{figure*}

\begin{figure*}[htb]
\centering
\includegraphics[width=0.99\textwidth]{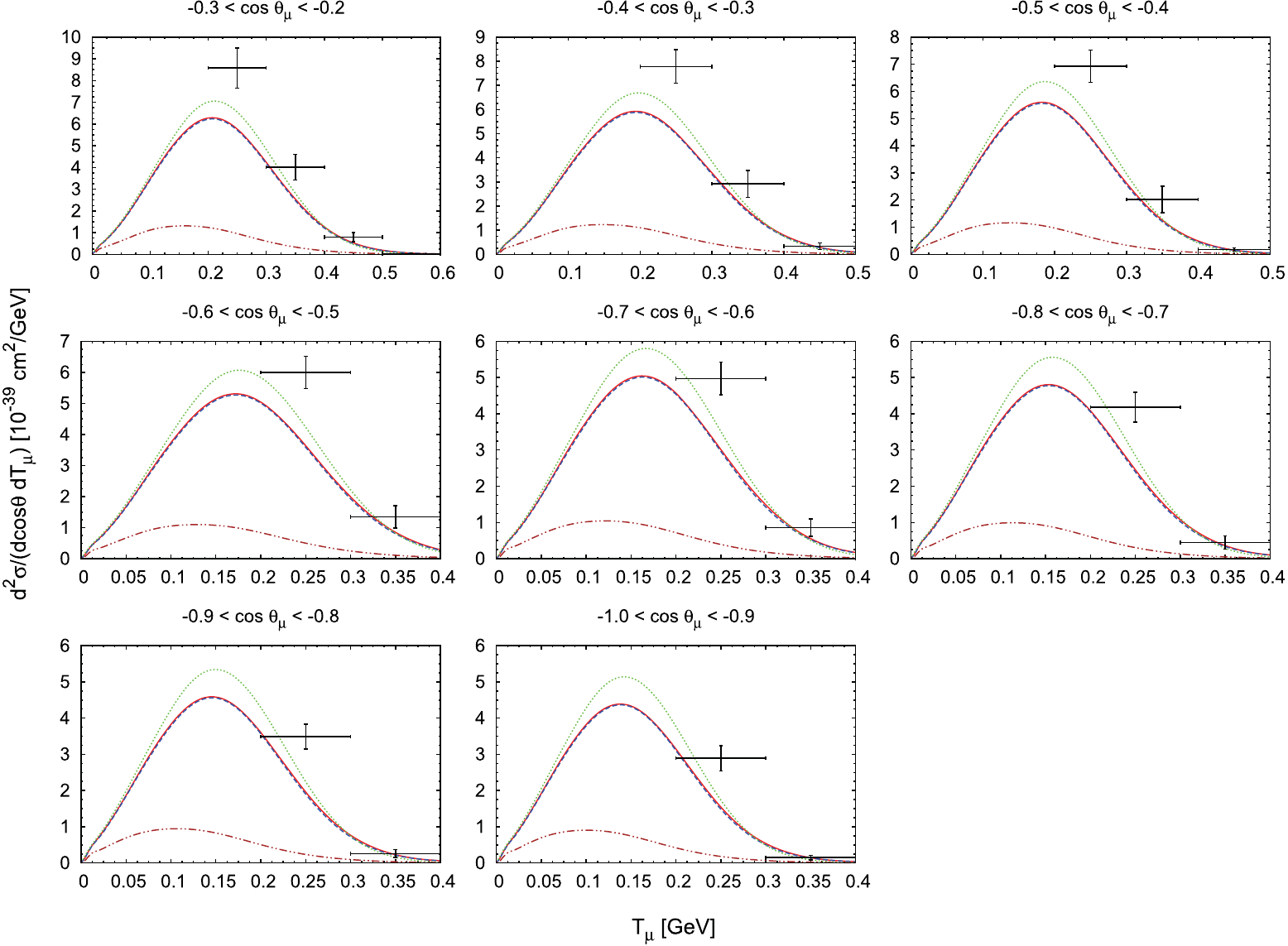}
\caption{(Color online) As for Fig.~\ref{nuMiniBooNE1}, but considering more backward kinematics. The data are from~\cite{miniboone}.
\label{nuMiniBooNE2}}
\end{figure*}

\begin{figure*}[htb]
\centering
\includegraphics[width=0.99\textwidth]{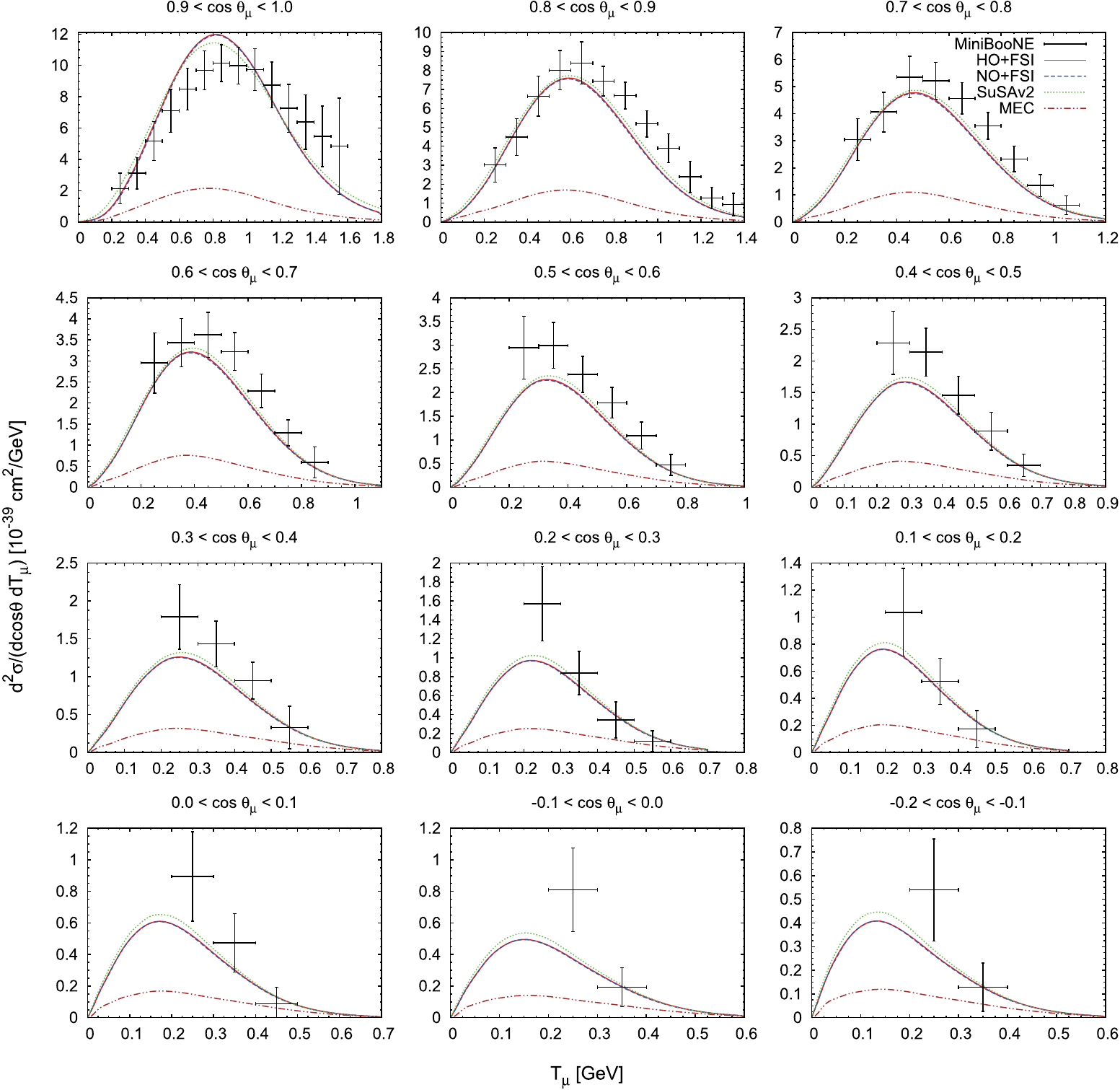}
\caption{(Color online) As for Fig.~\ref{nuMiniBooNE1}, but now for the $\overline{\nu}_\mu$ CCQE process on $^{12}$C. The data are from~\cite{miniboone-ant}.
\label{anuMiniBooNE1}}
\end{figure*}

\begin{figure*}[htb]
\centering
\includegraphics[width=0.99\textwidth]{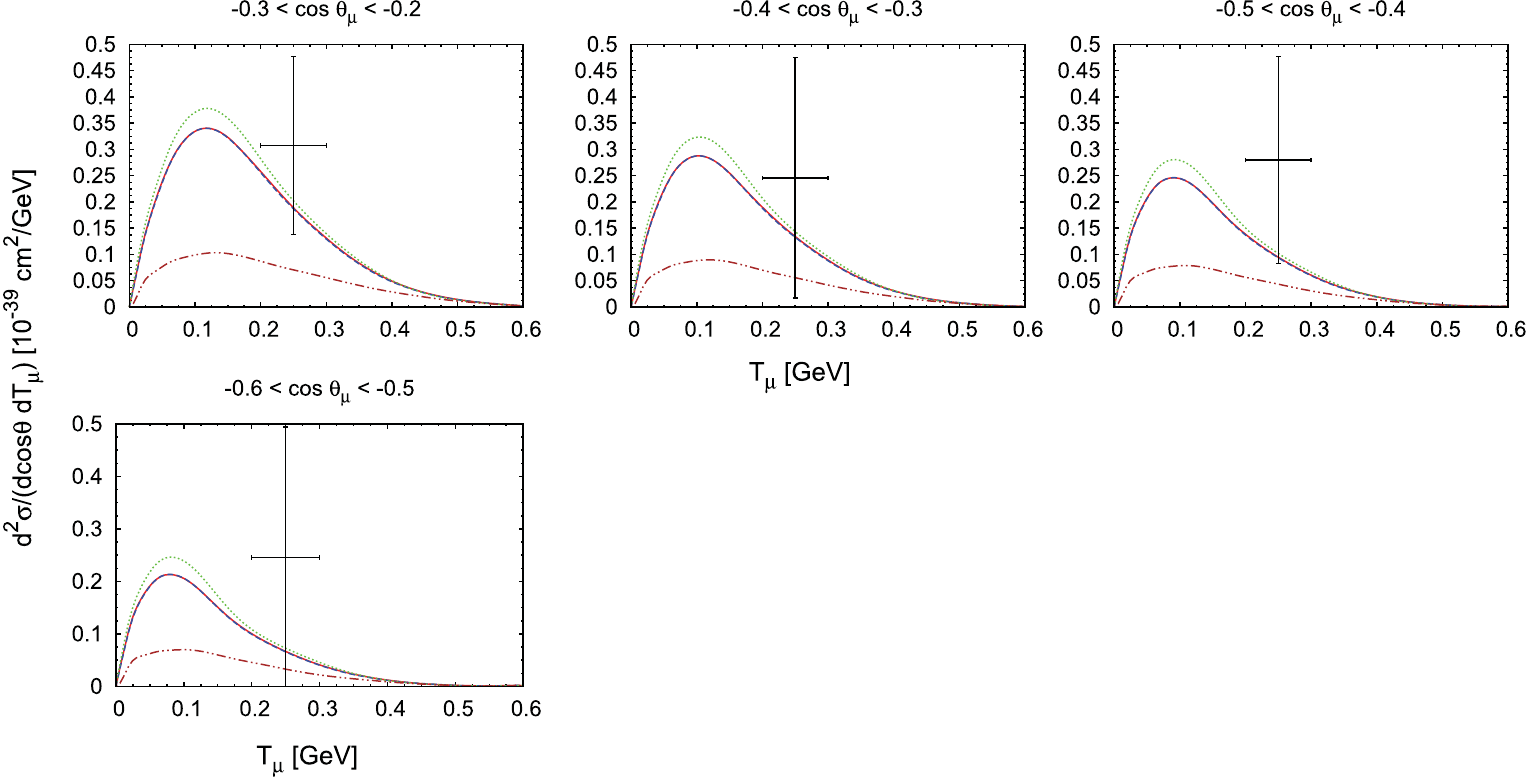}
\caption{(Color online) As for Fig.~\ref{anuMiniBooNE1}, but considering more backward kinematics. The data are from~\cite{miniboone-ant}.\label{anuMiniBooNE2}}
\end{figure*}

\begin{figure*}[htb]
\centering
\includegraphics[width=0.9\textwidth]{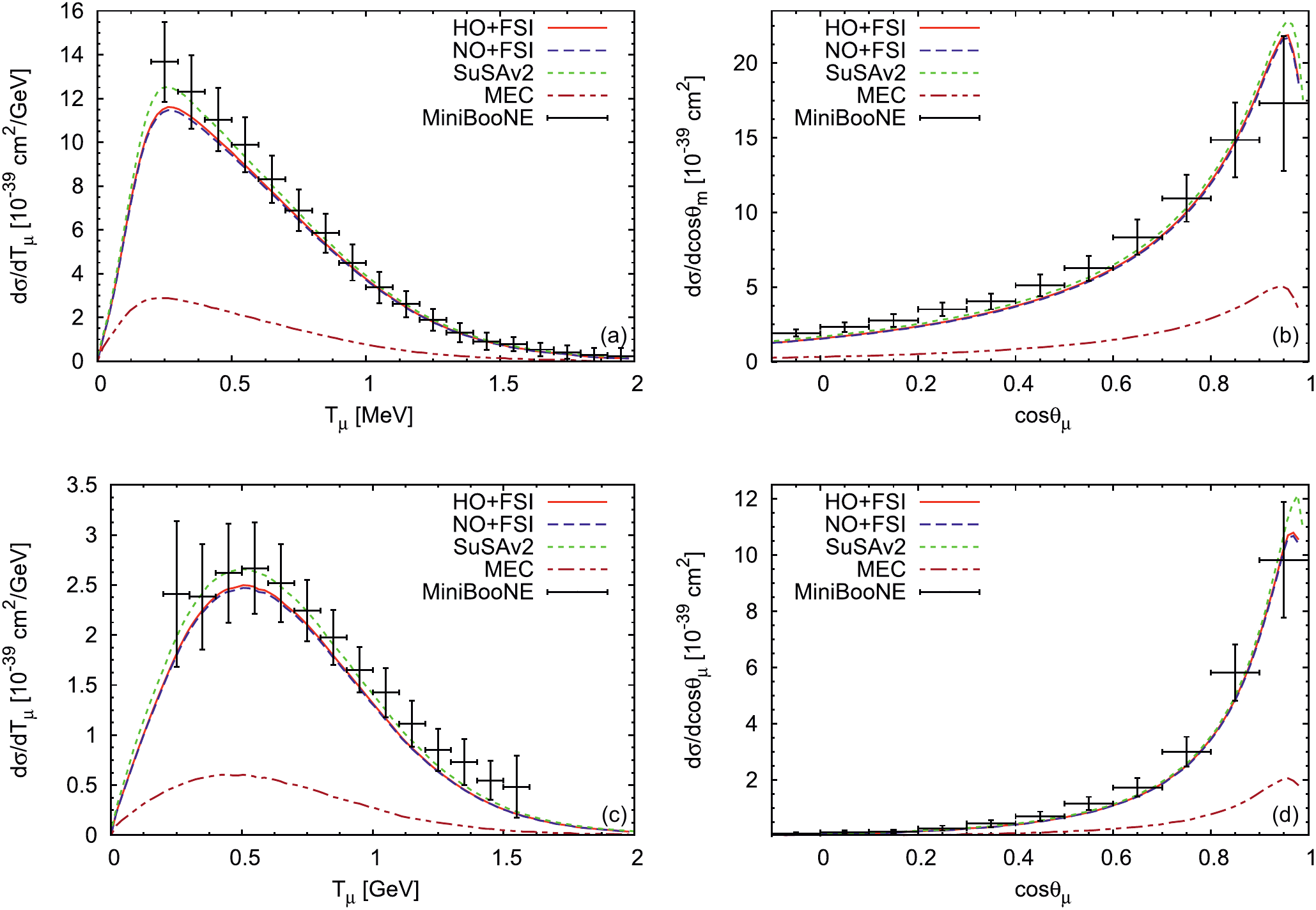}
\caption{(Color online) MiniBooNE flux-averaged CCQE $\nu_\mu$--$^{12}$C ($\overline{\nu}_\mu$--$^{12}$C) differential cross section per nucleon as a function of the muon kinetic energy [left panels -- (a) and (c)] and of the muon scattering angle [right panels -- (b) and (d)]. The top panels [(a) and (b)] correspond to neutrino cross sections and the bottom [(c) and (d)] ones to antineutrino reactions. The data are from~\cite{miniboone, miniboone-ant}.\label{diffMiniBooNE}}
\end{figure*}

\begin{figure*}[htb]
\centering
\includegraphics[width=8.5cm]{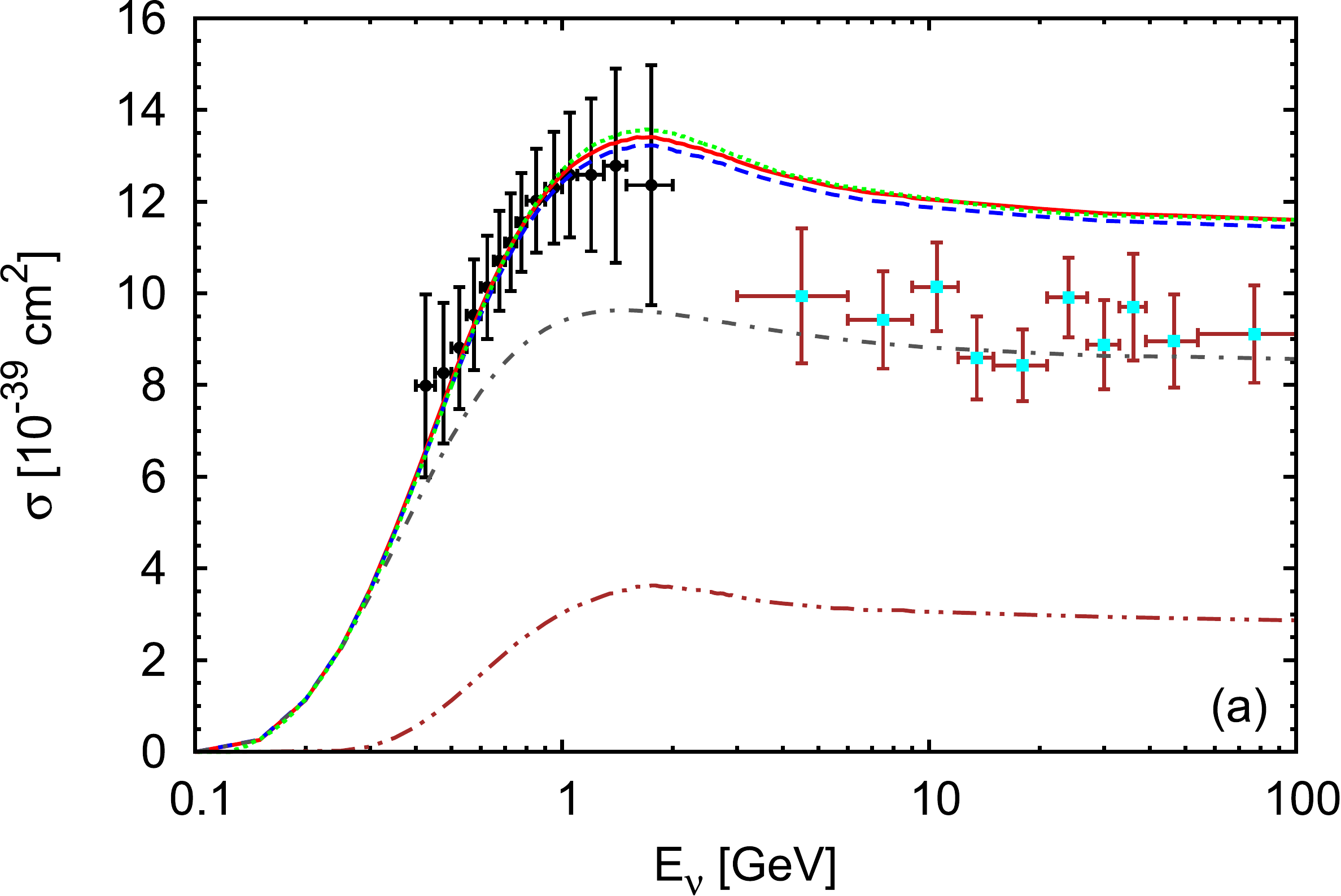}\hfill\includegraphics[width=8.5cm]{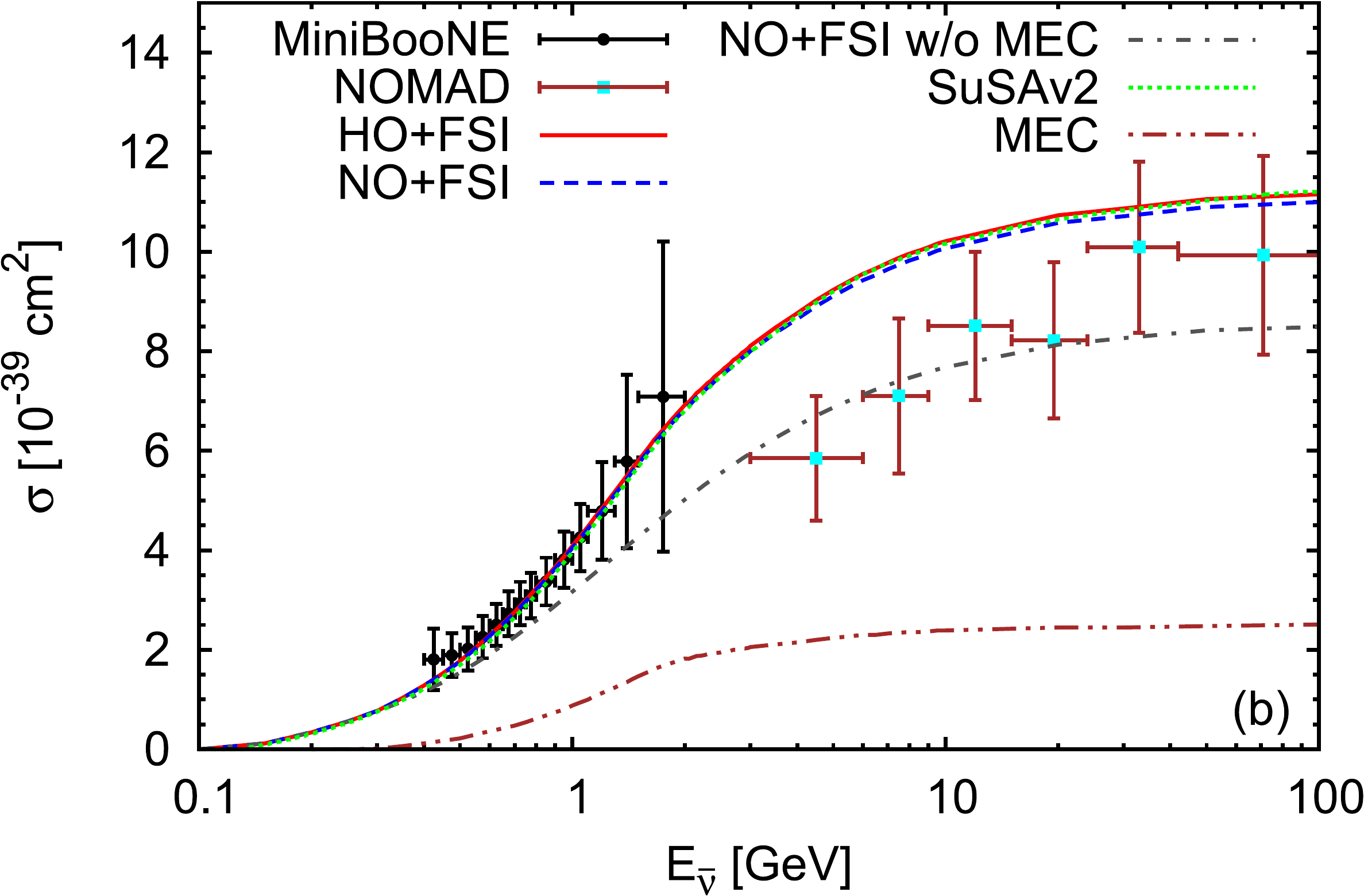}
\caption{(Color online) CCQE $\nu_\mu$--$^{12}$C ($\overline{\nu}_\mu$--$^{12}$C) total cross section per neutron (proton) as a function of the neutrino energy. The left panel (a) corresponds to neutrino cross sections and the right one (b) to antineutrino reactions. The data are from MiniBooNE~\cite{miniboone, miniboone-ant} and NOMAD~\cite{Lyubushkin:2009} experiments.\label{totMiniBooNE}}
\end{figure*}

In Figs.~\ref{nuMiniBooNE1}--\ref{anuMiniBooNE2} we show the double differential cross section averaged over the neutrino and antineutrino energy flux against the kinetic energy of the final muon. The data are taken from the MiniBooNE Collaboration~\cite{miniboone, miniboone-ant}. We represent a large variety of kinematical situations where each panel refers to results averaged over a particular muon angular bin.

We compare the data with the results obtained within the HO+FSI, NO+FSI, and SuSAv2 approaches, all of them including 2p--2h MEC, that are also presented separately. As already shown in~\cite{PhysRevD.94.093004}, notice the relevant role played by 2p-2h MEC contributions, of the order of $\sim$20-25$\%$ of the total response at the maximum. In the neutrino case (Figs.~\ref{nuMiniBooNE1} and~\ref{nuMiniBooNE2}) this relative strength is almost independent of the scattering angle, except for the most forward bin, $0.9 < \cos\theta < 1$, where the MEC contribution is $\sim$15$\%$; this angular bin, however, largely corresponds to very low excitation energies ($\omega <$ 50 MeV) and in this case completely different modeling, appropriate for the near-threshold regime, should be used. In the antineutrino case (Figs.~\ref{anuMiniBooNE1} and~\ref{anuMiniBooNE2}) the 2p-2h relative strength gets larger for backward scattering angles $(\cos\theta_\mu < -0.2)$. This is due to the fact that the antineutrino cross section involves a destructive interference between the $T$ and $T^\prime$ channels [see Eq.~(\ref{new33})] and is therefore more sensitive to nuclear effects.

Theoretical predictions including both the QE and the 2p-2h MEC contributions are in good accord with the data in most of the kinematical situations explored. Only at scattering angles approaching $90^\circ$ and above does one see a hint of a difference, although in these situations only a small number of data points with large uncertainties exist.

With regard to the comparison between the different models, we observe that HO+FSI and NO+FSI provide almost identical responses in all kinematical situations for neutrinos and antineutrinos: the inclusive cross section is not sensitive to the details of the spectral function. Compared with SuSAv2, some differences emerge whose magnitude depends on the scattering angle region explored. Whereas the SuSAv2 prediction is slightly smaller than the SF+FSI one at very forward kinematics (very small energy and momentum transfers), the reverse tends to occur as $\theta_\mu$ gets larger. Notice that at the most backward kinematics for neutrinos, the SuSAv2 results exceed by $\sim$15$\%$ those of the SF+FSI model at the maximum. Similar comments also apply to antineutrinos (Fig.~\ref{anuMiniBooNE1} and~\ref{anuMiniBooNE2}).

In Fig.~\ref{diffMiniBooNE} results are presented for the MiniBooNE flux averaged CCQE $\nu_\mu$--$^{12}$C ($\overline{\nu}_\mu$--$^{12}$C) single differential cross section per nucleon as a function of the muon kinetic energy [left panels -- (a) and (c)] and of the muon scattering angle [right panels -- (b) and (d)]. The top panels [(a) and (b)] correspond to neutrino cross sections and the bottom [(c) and (d)] ones to antineutrino reactions. Again, HO+FSI and NO+FSI lead to almost identical cross sections that differ from the SuSAv2 prediction by less than $\sim$5$\%-7\%$ at the maximum. The important contribution linked to the 2p-2h MEC (of the order of $\sim$20$\%-25\%$ of the total response) is clearly seen to be essential in order to describe the data.

To conclude this subsection, the results for the total flux-unfolded integrated cross sections per nucleon are given in Fig.~\ref{totMiniBooNE} and compared with the MiniBooNE~\cite{miniboone, miniboone-ant} and NOMAD~\cite{Lyubushkin:2009} data (up to $100$~GeV). In accordance with the previous discussion, 2p-2h MEC contributions are needed in order to reproduce MiniBooNE data.  On the contrary, the three theoretical models clearly overpredict the NOMAD data, these being more in agreement with the pure QE responses, for example is given NO+FSI without MEC result (see also ~\cite{PhysRevD.94.093004,Amaro:2013yna}). This result is consistent with the setup of NOMAD experiment that, unlike MiniBooNE, can select true QE, rather than ``QE-like'' events.  As observed, the discrepancy between the three theoretical predictions is very minor, but the role of the 2p-2h MEC is very significant at all neutrino energies, getting an almost constant value of the order of $\sim$30$\%-35\%$ compared with the pure QE contribution. The comparison of the theoretical calculations to the total flux-unfolded MiniBooNE cross sections data should be taken with caution because due to the multinucleon mechanism effects, the algorithm used to reconstruct the neutrino energy is not adequate when dealing with quasielastic-like events~\cite{PhysRevD.85.113008, PhysRevD.85.093012, Mosel:2012, PhysRevD.91.033005}.

Also, we would like to mention that the quasielastic data themselves have not directly been measured but have been deduced from so-called quasielastic-like data by subtracting a background of events in which pions are first produced, but then reabsorbed again. This background was determined from calculations with an event generator. Thus, the final QE + 2p-2h data invariably contain some model dependence~\cite{PhysRevC.87.014602}.

\subsection{MINER$\boldsymbol\nu$A\label{subsec32}}

\begin{figure}[htb]
\centering
\includegraphics[width=8.5cm]{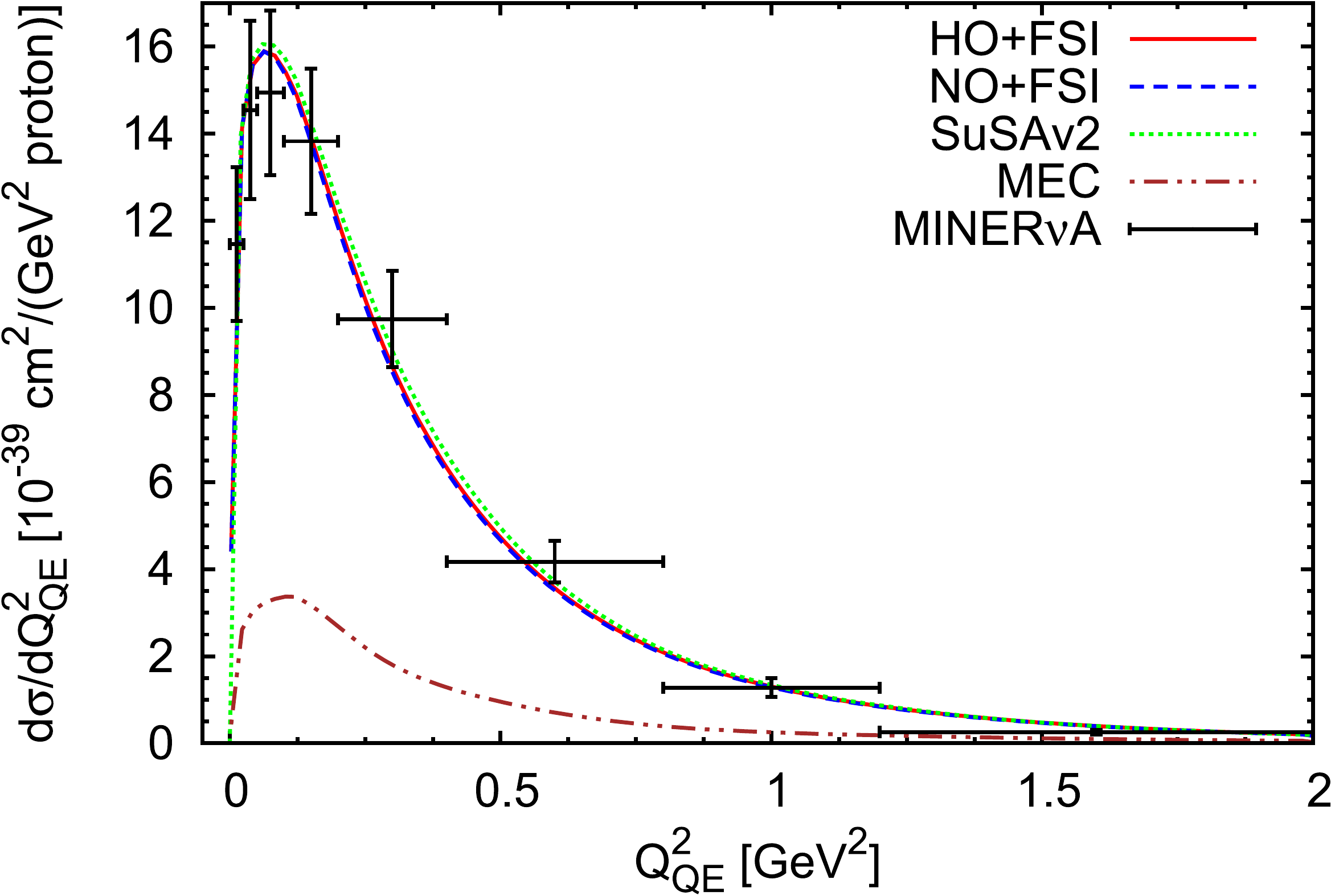}
\caption{(Color online) Flux-folded CCQE $\overline{\nu}_\mu$--CH scattering cross section per target proton as a function of $Q^2_\text{QE}$ and evaluated in the SuSAv2, HO+FSI, and NO+FSI approaches including MEC. The MINER$\nu$A data are from~\cite{cherylthesis}.\label{MINERvA}}
\end{figure}

The results in Fig.~\ref{MINERvA} correspond to the MINER$\nu$A flux averaged CCQE $\overline{\nu}_\mu$ differential cross section per nucleon as a function of the reconstructed four-momentum $Q^2_\text{QE}$, that is obtained in the same way as for the experiment, assuming an initial state nucleon at rest with a constant binding energy, $E_b$, set to $30$~MeV in the antineutrino case. The theoretical cross sections are flux-averaged using the new prediction of the NuMI flux~\cite{PhysRevD.94.092005} and are compared with $d\sigma/dQ^2_\text{QE}$, projected from the double-differential cross section~\cite{cherylthesis}.

As shown, the spread in the results ascribed to the three models used is minimal, of the order of $\sim$1$\%-2\%$. On the other hand, we note the excellent agreement between the theory and data once 2p-2h MEC effects ($\sim$20$\%-30\%$ of the total) are included. This significant contribution of the 2p-2h MEC effects is consistent with the results observed for MiniBooNE in spite of the very different muon antineutrino energy flux in the two experiments.

Recently, MINER$\nu$A collaboration has published new experimental data~\cite{PhysRevLett.119.082001}. In this work the cross sections are presented as a function of the four-momentum transfer $Q^2_p$, which in the case of CCQE scattering from a neutron at rest, can be calculated using the proton kinetic energy, $T_p$ alone. $Q^2_p$ is reconstructed based on the kinematics of the leading proton above tracking threshold ($p_{\text{proton}} > 450$~MeV/$c$). The analysis of the cross sections as a function of $Q^2_p$ is interesting for interpretation of the effects of FSI and this will be done in a new project.

\subsection{T2K\label{subsec33}}

\begin{figure*}[htb]
\centering
\includegraphics[width=0.99\textwidth]{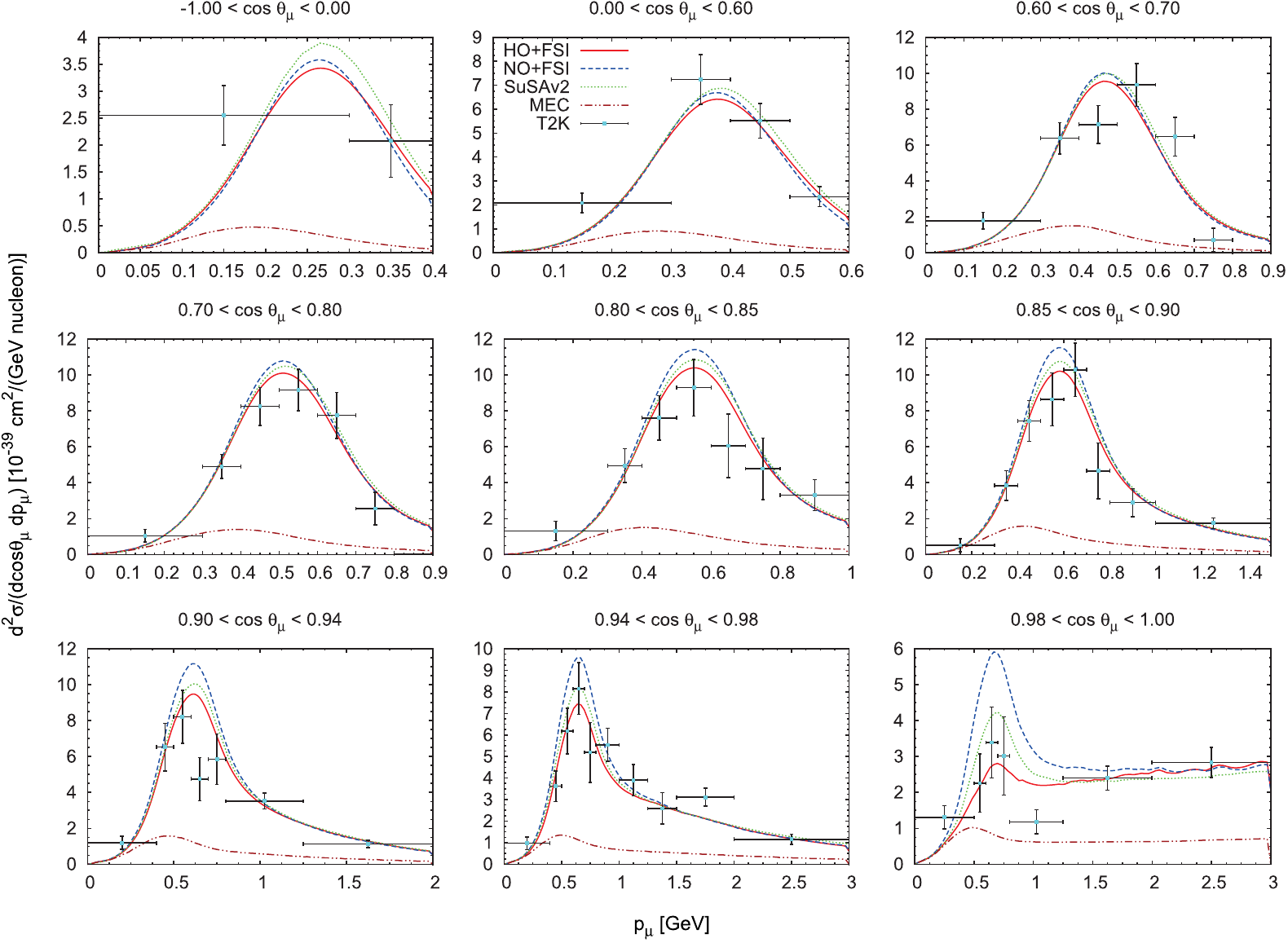}
\caption{(Color online) T2K flux-folded double differential cross section per target nucleon for the $\nu_\mu$ CCQE process on $^{12}$C displayed versus the $\mu^{-}$ momentum $p_{\mu}$ for various bins of $\cos \theta_\mu$ obtained within the SuSAv2, HO+FSI, and NO+FSI approaches including MEC. MEC results are shown also separately. The data are from~\cite{PhysRevD.93.112012}.\label{T2K}}
\end{figure*}

In Fig.~\ref{T2K} we present the flux-averaged double differential cross sections corresponding to the T2K experiment~\cite{PhysRevD.93.112012}. The graphs are plotted against the muon momentum, and each panel corresponds to a bin in the scattering angle. As in the previous cases, we show results obtained within the SuSAv2, HO+FSI, and NO+FSI approaches including MEC and also the separate contributions of the 2p-2h MEC. As already pointed out in \cite{PhysRevD.94.093004}, the narrower T2K flux, sharply peaked at about 0.7 GeV (see Fig.~\ref{fig:fluxes}), is the reason of the smaller contribution provided by the 2p-2h MEC (of the order of $\sim$10$\%$) as compared with the MiniBooNE and MINER$\nu$A results: in fact, the main contribution for the 2p-2h response comes from momentum transfers $q\sim$500 MeV/c, which are less important at T2K kinematics. Concerning the theoretical predictions, the two SF models produce almost identical cross sections that deviate from SuSAv2, particularly at backward kinematics (left-top panel) and very forward scattering (right-bottom panel). At backward angles this is consistent with the analysis presented for the MiniBooNE experiment.

In the particular case of the most forward scattering kinematics (bottom panel on the right), notice that SuSAv2 cross section at the maximum exceeds SF+FSI results by $\sim$30$\%-35\%$. However, the large error bands shown by T2K data do not allow us to discriminate between the different models, {\it i.e.,} neither between pure QE calculations nor global QE+2p-2h MEC results. Furthermore, notice that the cross section reaches an almost constant value, different from zero, as $p_\mu$ increases. This is in contrast with all remaining situations explored in the previous figures.

\begin{figure}[htb]
\centering
\includegraphics[width=8cm]{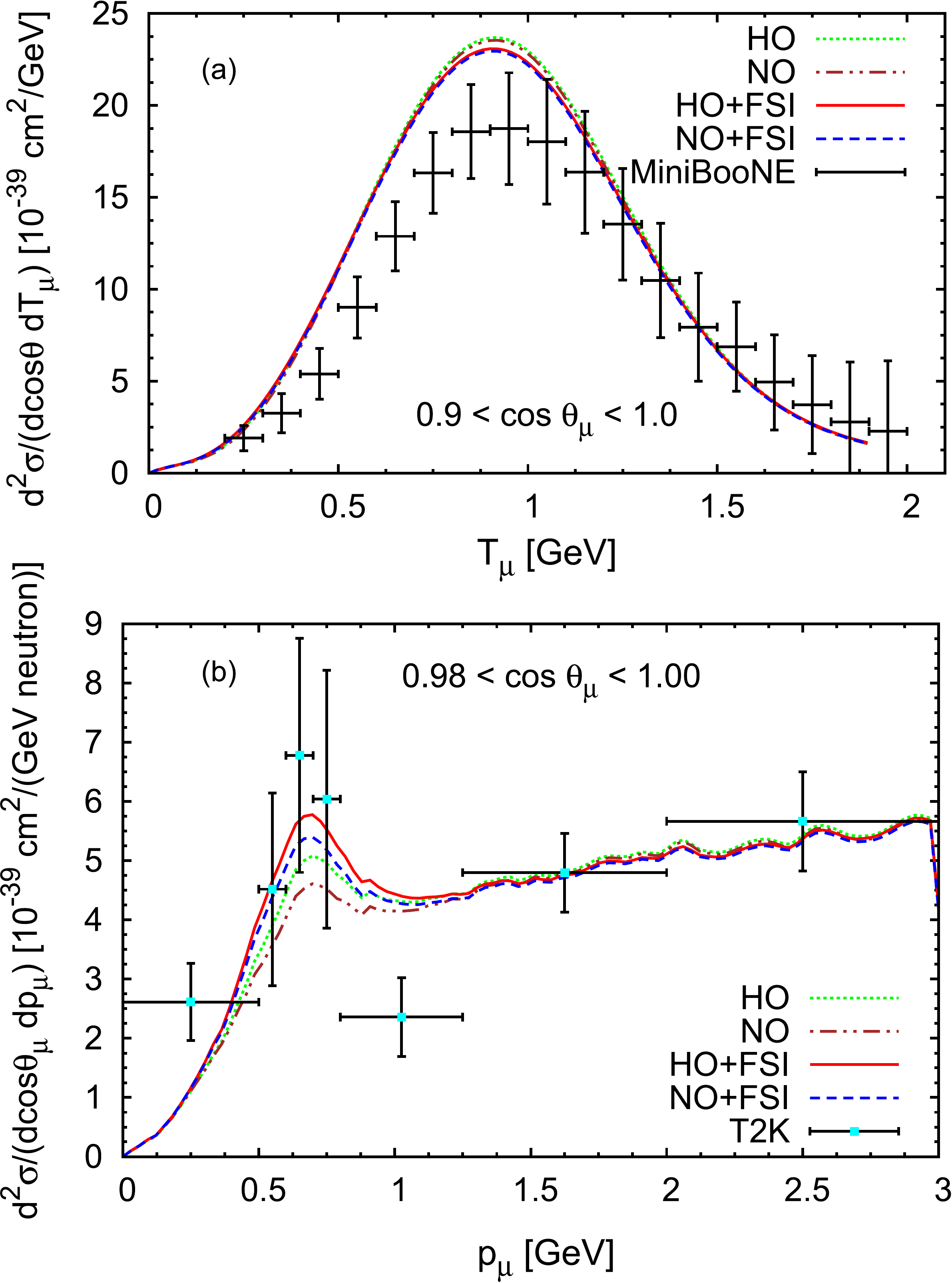}
\caption{(Color online) MiniBooNE flux-folded double differential cross section per target neutron for the $\nu_\mu$ CCQE process on $^{12}$C displayed versus the $\mu^{-}$ kinetic energy $T_\mu$ for $0.9<\cos \theta_\mu<1.0$ [top panel -- (a)] and T2K flux-folded double differential cross section per target neutron for the $\nu_\mu$ CCQE process on $^{12}$C displayed versus the $\mu^{-}$ momentum $p_{\mu}$ for $ 0.98<\cos \theta_\mu<1.00$ [bottom panel -- (b)] obtained within the HO, NO, HO+FSI, and NO+FSI approaches including MEC. The data are from~\cite{miniboone, PhysRevD.93.112012}.\label{nuwoFSI}}
\end{figure}
\begin{figure}[htb]
\centering
\includegraphics[width=8cm]{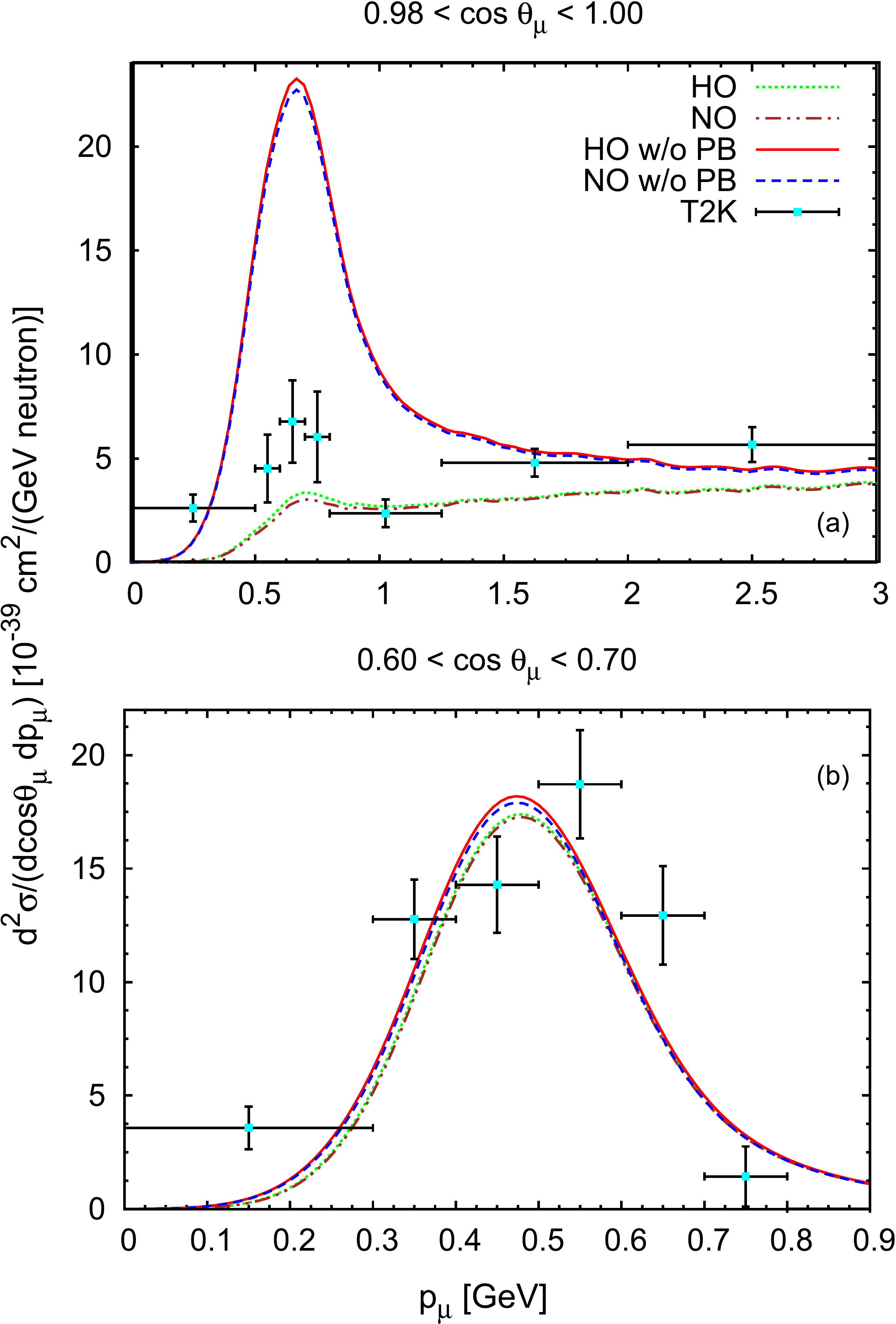}
\caption{(Color online) T2K flux-folded double differential cross section per target neutron for the $\nu_\mu$ CCQE process on $^{12}$C displayed versus the $\mu^{-}$ momentum $p_{\mu}$ for two bins of $ 0.98<\cos \theta_\mu<1.00$ and $ 0.60<\cos \theta_\mu<0.70$ obtained within the HO+FSI and NO+FSI approaches with and without PB effects. Data are from~\cite{PhysRevD.93.112012}.\label{nuwoFSIwoPB}}
\end{figure}

Before concluding, we provide some closer inspection of the results at the most forward scattering angles, which represent a very delicate and model-dependent kinematical situation.
In Fig.~\ref{nuwoFSI} we present the MiniBooNE [panel (a)] and T2K [panel (b)] flux-folded double differential cross sections for the $\nu_\mu$ CCQE process on $^{12}$C for very forward angles obtained within HO and NO approaches with and without FSI. As was shown in ~\cite{PhysRevC.89.014607} the effects of FSI on the total cross section consist of an increase of about 2\% using HO and NO spectral functions, almost independently of the neutrino energy. It can be seen in Fig.~\ref{nuwoFSI} that the effects of FSI on the double differential cross section is a bit larger for the most forward angles; it is about 10\% in the case of T2K experiment and about 2--4\% at the peak for the MiniBooNE experiment. Finally, in Fig.~\ref{nuwoFSIwoPB} we show the T2K flux-folded double differential cross section per target neutron versus $p_{\mu}$ for two bins of $\theta_\mu$ [$0.98<\cos \theta_\mu<1.00$ (a) and $0.60<\cos \theta_\mu<0.70$ (b)] obtained within the HO and NO approaches with and without PB effects. The PB effects play a significant role in the case of the most forward angles [Fig.~\ref{nuwoFSIwoPB}(a)] and decrease as the muon angle $\theta_\mu$ grows [Fig.~\ref{nuwoFSIwoPB}(b)]. These results are in line with what has been shown in~\cite{Megias:2017cuh} using a different spectral function model~\cite{Benhar:1994hw} and confirms the fact that the low energy transfer region is crucially important in the description of forward scattering data.

\section{Conclusions\label{sec:conclusions}}

This work extends our previous studies of CCQE neutrino-nucleus scattering processes that are of interest for neutrino (antineutrino) oscillation experiments. Here we focus on models based on the use of  two spectral functions, one of them including \emph{NN} short-range correlations through the Jastrow method and, for a comparison, another without them. Effects of final-state interactions are also incorporated  by using an optical potential. These calculations, based on the impulse approximation, are complemented with the contributions given by two-body weak meson exchange currents, giving rise to two-particle two-hole excitations.

The model is applied to three different experiments, MiniBooNE, MINER$\nu$A and T2K, spanning a wide range of neutrino energies and all scattering angle values, from forward to backward kinematics.

These new predictions are compared with the systematic analysis presented in~\cite{PhysRevD.94.093004} based on the SuSAv2+MEC approach. We find that the spectral function based models (HO+FSI, NO+FSI) lead to results that are very close to the SuSAv2-MEC predictions. Only at the most forward and most backward angles do the differences become larger, being at most of the order of $\sim$10$\%-12\%$. In the case of single-differential cross sections and, particularly, for the total flux-unfolded integrated cross section, these model differences become very minor, almost negligible. This is in contrast with the contribution ascribed to the 2p-2h MEC effects that can be even larger than $\sim$30$\%-35\%$ compared with the pure QE responses. This proves without ambiguity the essential role played by 2p-2h MEC in providing a successful description of neutrino (antineutrino)-nucleus scattering data for different experiments and a very wide range of kinematical situations.

An interesting outcome of the present study is that the results obtained with the NO spectral function, which accounts for \emph{NN} short-range Jastrow correlations, are almost identical to those obtained with the uncorrelated HO spectral function, thus indicating that the role played by this type of correlations is very minor for the observables analyzed in this study. Comparisons with a different spectral function model~\cite{Megias:2017cuh} also show that inclusive reactions -- where only the outgoing lepton is measured -- are not sensitive to the detailed description of the nuclear initial state. Nevertheless, the results in this work can be seen as a test of the reliability of the present spectral function based models. They compare extremely well with the SuSAv2 approach, based on the phenomenology of electron scattering data, although they fail in reproducing neutrino (antineutrino) scattering data unless ingredients beyond the impulse approximation are incorporated. The present study gives us confidence in extending the use of these models to other processes, such as semi-inclusive CC$\nu$ reactions and neutral current processes.

\begin{acknowledgments}

This work was partially supported by the Bulgarian National Science Fund under Contracts Nos. DFNI-T02/19, DFNI-E02/6, and DNTS/Russia~01/3,  by the Russian Foundation for Basic Research grant No.~17-52-18057-bolg-a, by the Spanish Ministerio de Economia y Competitividad and ERDF (European Regional Development Fund) under contracts FIS2014-59386-P, FIS2014-53448-C2-1, FIS2017-88410-P, FIS2017-85053-C2-1-P, FPA2015-65035-P, by the Junta de Andalucia (grants No. FQM-225, FQM160), by the INFN under project MANYBODY, by the University of Turin under contract BARM-RILO-17, and part (TWD) by the U.S. Department of Energy under cooperative agreement DE-FC02-94ER40818. GDM acknowledges support from a Junta de Andalucia fellowship (FQM7632, Proyectos de Excelencia 2011). MBB acknowledges support from the ``Emilie du Ch\^atelet" programme of the P2IO LabEx (ANR-10-LABX-0038).

\end{acknowledgments}

%%%%%%%%%%
%\bibliographystyle{plain}
\bibliography{biblio}

\begin{thebibliography}{78}%
\makeatletter
\providecommand \@ifxundefined [1]{%
 \@ifx{#1\undefined}
}%
\providecommand \@ifnum [1]{%
 \ifnum #1\expandafter \@firstoftwo
 \else \expandafter \@secondoftwo
 \fi
}%
\providecommand \@ifx [1]{%
 \ifx #1\expandafter \@firstoftwo
 \else \expandafter \@secondoftwo
 \fi
}%
\providecommand \natexlab [1]{#1}%
\providecommand \enquote  [1]{``#1''}%
\providecommand \bibnamefont  [1]{#1}%
\providecommand \bibfnamefont [1]{#1}%
\providecommand \citenamefont [1]{#1}%
\providecommand \href@noop [0]{\@secondoftwo}%
\providecommand \href [0]{\begingroup \@sanitize@url \@href}%
\providecommand \@href[1]{\@@startlink{#1}\@@href}%
\providecommand \@@href[1]{\endgroup#1\@@endlink}%
\providecommand \@sanitize@url [0]{\catcode `\\12\catcode `\$12\catcode
  `\&12\catcode `\#12\catcode `\^12\catcode `\_12\catcode `\%12\relax}%
\providecommand \@@startlink[1]{}%
\providecommand \@@endlink[0]{}%
\providecommand \url  [0]{\begingroup\@sanitize@url \@url }%
\providecommand \@url [1]{\endgroup\@href {#1}{\urlprefix }}%
\providecommand \urlprefix  [0]{URL }%
\providecommand \Eprint [0]{\href }%
\providecommand \doibase [0]{http://dx.doi.org/}%
\providecommand \selectlanguage [0]{\@gobble}%
\providecommand \bibinfo  [0]{\@secondoftwo}%
\providecommand \bibfield  [0]{\@secondoftwo}%
\providecommand \translation [1]{[#1]}%
\providecommand \BibitemOpen [0]{}%
\providecommand \bibitemStop [0]{}%
\providecommand \bibitemNoStop [0]{.\EOS\space}%
\providecommand \EOS [0]{\spacefactor3000\relax}%
\providecommand \BibitemShut  [1]{\csname bibitem#1\endcsname}%
\let\auto@bib@innerbib\@empty
%</preamble>
\bibitem [{\citenamefont {Alvarez-Ruso}\ \emph {et~al.}(2018)\citenamefont
  {Alvarez-Ruso} \emph {et~al.}}]{Alvarez-Ruso:2017oui}%
  \BibitemOpen
  \bibfield  {author} {\bibinfo {author} {\bibfnamefont {L.}~\bibnamefont
  {Alvarez-Ruso}} \emph {et~al.},\ }\href {\doibase 10.1016/j.ppnp.2018.01.006}
  {\bibfield  {journal} {\bibinfo  {journal} {Prog. Part. Nucl. Phys.}\
  }\textbf {\bibinfo {volume} {100}},\ \bibinfo {pages} {1} (\bibinfo {year}
  {2018})}\BibitemShut {NoStop}%
\bibitem [{\citenamefont {Megias}\ \emph
  {et~al.}(2016{\natexlab{a}})\citenamefont {Megias}, \citenamefont {Amaro},
  \citenamefont {Barbaro}, \citenamefont {Caballero}, \citenamefont
  {Donnelly},\ and\ \citenamefont {Simo}}]{PhysRevD.94.093004}%
  \BibitemOpen
  \bibfield  {author} {\bibinfo {author} {\bibfnamefont {G.~D.}\ \bibnamefont
  {Megias}}, \bibinfo {author} {\bibfnamefont {J.~E.}\ \bibnamefont {Amaro}},
  \bibinfo {author} {\bibfnamefont {M.~B.}\ \bibnamefont {Barbaro}}, \bibinfo
  {author} {\bibfnamefont {J.~A.}\ \bibnamefont {Caballero}}, \bibinfo {author}
  {\bibfnamefont {T.~W.}\ \bibnamefont {Donnelly}}, \ and\ \bibinfo {author}
  {\bibfnamefont {I.~R.}\ \bibnamefont {Simo}},\ }\href {\doibase
  10.1103/PhysRevD.94.093004} {\bibfield  {journal} {\bibinfo  {journal} {Phys.
  Rev. D}\ }\textbf {\bibinfo {volume} {94}},\ \bibinfo {pages} {093004}
  (\bibinfo {year} {2016}{\natexlab{a}})}\BibitemShut {NoStop}%
\bibitem [{\citenamefont {Amaro}\ \emph {et~al.}(2005)\citenamefont {Amaro},
  \citenamefont {Barbaro}, \citenamefont {Caballero}, \citenamefont {Donnelly},
  \citenamefont {Molinari},\ and\ \citenamefont {Sick}}]{Amaro:2004bs}%
  \BibitemOpen
  \bibfield  {author} {\bibinfo {author} {\bibfnamefont {J.~E.}\ \bibnamefont
  {Amaro}}, \bibinfo {author} {\bibfnamefont {M.~B.}\ \bibnamefont {Barbaro}},
  \bibinfo {author} {\bibfnamefont {J.~A.}\ \bibnamefont {Caballero}}, \bibinfo
  {author} {\bibfnamefont {T.~W.}\ \bibnamefont {Donnelly}}, \bibinfo {author}
  {\bibfnamefont {A.}~\bibnamefont {Molinari}}, \ and\ \bibinfo {author}
  {\bibfnamefont {I.}~\bibnamefont {Sick}},\ }\href {\doibase
  10.1103/PhysRevC.71.015501} {\bibfield  {journal} {\bibinfo  {journal} {Phys.
  Rev. C}\ }\textbf {\bibinfo {volume} {71}},\ \bibinfo {pages} {015501}
  (\bibinfo {year} {2005})}\BibitemShut {NoStop}%
\bibitem [{\citenamefont {Amaro}\ \emph {et~al.}(2007)\citenamefont {Amaro},
  \citenamefont {Barbaro}, \citenamefont {Caballero},\ and\ \citenamefont
  {Donnelly}}]{Amaro:2006tf}%
  \BibitemOpen
  \bibfield  {author} {\bibinfo {author} {\bibfnamefont {J.~E.}\ \bibnamefont
  {Amaro}}, \bibinfo {author} {\bibfnamefont {M.~B.}\ \bibnamefont {Barbaro}},
  \bibinfo {author} {\bibfnamefont {J.~A.}\ \bibnamefont {Caballero}}, \ and\
  \bibinfo {author} {\bibfnamefont {T.~W.}\ \bibnamefont {Donnelly}},\ }\href
  {\doibase 10.1103/PhysRevLett.98.242501} {\bibfield  {journal} {\bibinfo
  {journal} {Phys. Rev. Lett.}\ }\textbf {\bibinfo {volume} {98}},\ \bibinfo
  {pages} {242501} (\bibinfo {year} {2007})}\BibitemShut {NoStop}%
\bibitem [{\citenamefont {Day}\ \emph {et~al.}(1990)\citenamefont {Day},
  \citenamefont {McCarthy}, \citenamefont {Donnelly},\ and\ \citenamefont
  {Sick}}]{doi:10.1146/annurev.ns.40.120190.002041}%
  \BibitemOpen
  \bibfield  {author} {\bibinfo {author} {\bibfnamefont {D.~B.}\ \bibnamefont
  {Day}}, \bibinfo {author} {\bibfnamefont {J.~S.}\ \bibnamefont {McCarthy}},
  \bibinfo {author} {\bibfnamefont {T.~W.}\ \bibnamefont {Donnelly}}, \ and\
  \bibinfo {author} {\bibfnamefont {I.}~\bibnamefont {Sick}},\ }\href {\doibase
  10.1146/annurev.ns.40.120190.002041} {\bibfield  {journal} {\bibinfo
  {journal} {Annu. Rev. Nucl. Part. Sci.}\ }\textbf {\bibinfo {volume} {40}},\
  \bibinfo {pages} {357} (\bibinfo {year} {1990})}\BibitemShut {NoStop}%
\bibitem [{\citenamefont {Sick}\ \emph {et~al.}(1980)\citenamefont {Sick},
  \citenamefont {Day},\ and\ \citenamefont {McCarthy}}]{PhysRevLett.45.871}%
  \BibitemOpen
  \bibfield  {author} {\bibinfo {author} {\bibfnamefont {I.}~\bibnamefont
  {Sick}}, \bibinfo {author} {\bibfnamefont {D.}~\bibnamefont {Day}}, \ and\
  \bibinfo {author} {\bibfnamefont {J.~S.}\ \bibnamefont {McCarthy}},\ }\href
  {\doibase 10.1103/PhysRevLett.45.871} {\bibfield  {journal} {\bibinfo
  {journal} {Phys. Rev. Lett.}\ }\textbf {\bibinfo {volume} {45}},\ \bibinfo
  {pages} {871} (\bibinfo {year} {1980})}\BibitemShut {NoStop}%
\bibitem [{\citenamefont {Ciofi~degli Atti}\ \emph {et~al.}(1987)\citenamefont
  {Ciofi~degli Atti}, \citenamefont {Pace},\ and\ \citenamefont
  {Salm\`e}}]{PhysRevC.36.1208}%
  \BibitemOpen
  \bibfield  {author} {\bibinfo {author} {\bibfnamefont {C.}~\bibnamefont
  {Ciofi~degli Atti}}, \bibinfo {author} {\bibfnamefont {E.}~\bibnamefont
  {Pace}}, \ and\ \bibinfo {author} {\bibfnamefont {G.}~\bibnamefont
  {Salm\`e}},\ }\href {\doibase 10.1103/PhysRevC.36.1208} {\bibfield  {journal}
  {\bibinfo  {journal} {Phys. Rev. C}\ }\textbf {\bibinfo {volume} {36}},\
  \bibinfo {pages} {1208} (\bibinfo {year} {1987})}\BibitemShut {NoStop}%
\bibitem [{\citenamefont {Ciofi~degli Atti}\ \emph {et~al.}(1989)\citenamefont
  {Ciofi~degli Atti}, \citenamefont {Pace},\ and\ \citenamefont
  {Salm\`e}}]{PhysRevC.39.259}%
  \BibitemOpen
  \bibfield  {author} {\bibinfo {author} {\bibfnamefont {C.}~\bibnamefont
  {Ciofi~degli Atti}}, \bibinfo {author} {\bibfnamefont {E.}~\bibnamefont
  {Pace}}, \ and\ \bibinfo {author} {\bibfnamefont {G.}~\bibnamefont
  {Salm\`e}},\ }\href {\doibase 10.1103/PhysRevC.39.259} {\bibfield  {journal}
  {\bibinfo  {journal} {Phys. Rev. C}\ }\textbf {\bibinfo {volume} {39}},\
  \bibinfo {pages} {259} (\bibinfo {year} {1989})}\BibitemShut {NoStop}%
\bibitem [{\citenamefont {Ciofi~degli Atti}\ \emph {et~al.}(1991)\citenamefont
  {Ciofi~degli Atti}, \citenamefont {Pace},\ and\ \citenamefont
  {Salm\`e}}]{PhysRevC.43.1155}%
  \BibitemOpen
  \bibfield  {author} {\bibinfo {author} {\bibfnamefont {C.}~\bibnamefont
  {Ciofi~degli Atti}}, \bibinfo {author} {\bibfnamefont {E.}~\bibnamefont
  {Pace}}, \ and\ \bibinfo {author} {\bibfnamefont {G.}~\bibnamefont
  {Salm\`e}},\ }\href {\doibase 10.1103/PhysRevC.43.1155} {\bibfield  {journal}
  {\bibinfo  {journal} {Phys. Rev. C}\ }\textbf {\bibinfo {volume} {43}},\
  \bibinfo {pages} {1155} (\bibinfo {year} {1991})}\BibitemShut {NoStop}%
\bibitem [{\citenamefont {Ciofi~degli Atti}\ \emph {et~al.}(1992)\citenamefont
  {Ciofi~degli Atti}, \citenamefont {Day},\ and\ \citenamefont
  {Liuti}}]{PhysRevC.46.1045}%
  \BibitemOpen
  \bibfield  {author} {\bibinfo {author} {\bibfnamefont {C.}~\bibnamefont
  {Ciofi~degli Atti}}, \bibinfo {author} {\bibfnamefont {D.~B.}\ \bibnamefont
  {Day}}, \ and\ \bibinfo {author} {\bibfnamefont {S.}~\bibnamefont {Liuti}},\
  }\href {\doibase 10.1103/PhysRevC.46.1045} {\bibfield  {journal} {\bibinfo
  {journal} {Phys. Rev. C}\ }\textbf {\bibinfo {volume} {46}},\ \bibinfo
  {pages} {1045} (\bibinfo {year} {1992})}\BibitemShut {NoStop}%
\bibitem [{\citenamefont {Ciofi~degli Atti}\ and\ \citenamefont
  {Simula}(1996)}]{PhysRevC.53.1689}%
  \BibitemOpen
  \bibfield  {author} {\bibinfo {author} {\bibfnamefont {C.}~\bibnamefont
  {Ciofi~degli Atti}}\ and\ \bibinfo {author} {\bibfnamefont {S.}~\bibnamefont
  {Simula}},\ }\href {\doibase 10.1103/PhysRevC.53.1689} {\bibfield  {journal}
  {\bibinfo  {journal} {Phys. Rev. C}\ }\textbf {\bibinfo {volume} {53}},\
  \bibinfo {pages} {1689} (\bibinfo {year} {1996})}\BibitemShut {NoStop}%
\bibitem [{\citenamefont {degli Atti}\ and\ \citenamefont
  {West}(1999)}]{degliAtti1999447}%
  \BibitemOpen
  \bibfield  {author} {\bibinfo {author} {\bibfnamefont {C.~C.}\ \bibnamefont
  {degli Atti}}\ and\ \bibinfo {author} {\bibfnamefont {G.~B.}\ \bibnamefont
  {West}},\ }\href {\doibase http://dx.doi.org/10.1016/S0370-2693(99)00599-7}
  {\bibfield  {journal} {\bibinfo  {journal} {Physics Letters B}\ }\textbf
  {\bibinfo {volume} {458}},\ \bibinfo {pages} {447 } (\bibinfo {year}
  {1999})}\BibitemShut {NoStop}%
\bibitem [{\citenamefont {Alberico}\ \emph {et~al.}(1988)\citenamefont
  {Alberico}, \citenamefont {Molinari}, \citenamefont {Donnelly}, \citenamefont
  {Kronenberg},\ and\ \citenamefont {Van~Orden}}]{PhysRevC.38.1801}%
  \BibitemOpen
  \bibfield  {author} {\bibinfo {author} {\bibfnamefont {W.~M.}\ \bibnamefont
  {Alberico}}, \bibinfo {author} {\bibfnamefont {A.}~\bibnamefont {Molinari}},
  \bibinfo {author} {\bibfnamefont {T.~W.}\ \bibnamefont {Donnelly}}, \bibinfo
  {author} {\bibfnamefont {E.~L.}\ \bibnamefont {Kronenberg}}, \ and\ \bibinfo
  {author} {\bibfnamefont {J.~W.}\ \bibnamefont {Van~Orden}},\ }\href {\doibase
  10.1103/PhysRevC.38.1801} {\bibfield  {journal} {\bibinfo  {journal} {Phys.
  Rev. C}\ }\textbf {\bibinfo {volume} {38}},\ \bibinfo {pages} {1801}
  (\bibinfo {year} {1988})}\BibitemShut {NoStop}%
\bibitem [{\citenamefont {Barbaro}\ \emph {et~al.}(1998)\citenamefont
  {Barbaro}, \citenamefont {Cenni}, \citenamefont {Pace}, \citenamefont
  {Donnelly},\ and\ \citenamefont {Molinari}}]{Barbaro1998137}%
  \BibitemOpen
  \bibfield  {author} {\bibinfo {author} {\bibfnamefont {M.}~\bibnamefont
  {Barbaro}}, \bibinfo {author} {\bibfnamefont {R.}~\bibnamefont {Cenni}},
  \bibinfo {author} {\bibfnamefont {A.~D.}\ \bibnamefont {Pace}}, \bibinfo
  {author} {\bibfnamefont {T.}~\bibnamefont {Donnelly}}, \ and\ \bibinfo
  {author} {\bibfnamefont {A.}~\bibnamefont {Molinari}},\ }\href {\doibase
  http://dx.doi.org/10.1016/S0375-9474(98)00443-6} {\bibfield  {journal}
  {\bibinfo  {journal} {Nuclear Physics A}\ }\textbf {\bibinfo {volume}
  {643}},\ \bibinfo {pages} {137 } (\bibinfo {year} {1998})}\BibitemShut
  {NoStop}%
\bibitem [{\citenamefont {Donnelly}\ and\ \citenamefont
  {Sick}(1999{\natexlab{a}})}]{PhysRevLett.82.3212}%
  \BibitemOpen
  \bibfield  {author} {\bibinfo {author} {\bibfnamefont {T.~W.}\ \bibnamefont
  {Donnelly}}\ and\ \bibinfo {author} {\bibfnamefont {I.}~\bibnamefont
  {Sick}},\ }\href {\doibase 10.1103/PhysRevLett.82.3212} {\bibfield  {journal}
  {\bibinfo  {journal} {Phys. Rev. Lett.}\ }\textbf {\bibinfo {volume} {82}},\
  \bibinfo {pages} {3212} (\bibinfo {year} {1999}{\natexlab{a}})}\BibitemShut
  {NoStop}%
\bibitem [{\citenamefont {Donnelly}\ and\ \citenamefont
  {Sick}(1999{\natexlab{b}})}]{PhysRevC.60.065502}%
  \BibitemOpen
  \bibfield  {author} {\bibinfo {author} {\bibfnamefont {T.~W.}\ \bibnamefont
  {Donnelly}}\ and\ \bibinfo {author} {\bibfnamefont {I.}~\bibnamefont
  {Sick}},\ }\href {\doibase 10.1103/PhysRevC.60.065502} {\bibfield  {journal}
  {\bibinfo  {journal} {Phys. Rev. C}\ }\textbf {\bibinfo {volume} {60}},\
  \bibinfo {pages} {065502} (\bibinfo {year} {1999}{\natexlab{b}})}\BibitemShut
  {NoStop}%
\bibitem [{\citenamefont {Maieron}\ \emph {et~al.}(2002)\citenamefont
  {Maieron}, \citenamefont {Donnelly},\ and\ \citenamefont
  {Sick}}]{PhysRevC.65.025502}%
  \BibitemOpen
  \bibfield  {author} {\bibinfo {author} {\bibfnamefont {C.}~\bibnamefont
  {Maieron}}, \bibinfo {author} {\bibfnamefont {T.~W.}\ \bibnamefont
  {Donnelly}}, \ and\ \bibinfo {author} {\bibfnamefont {I.}~\bibnamefont
  {Sick}},\ }\href {\doibase 10.1103/PhysRevC.65.025502} {\bibfield  {journal}
  {\bibinfo  {journal} {Phys. Rev. C}\ }\textbf {\bibinfo {volume} {65}},\
  \bibinfo {pages} {025502} (\bibinfo {year} {2002})}\BibitemShut {NoStop}%
\bibitem [{\citenamefont {Barbaro}\ \emph {et~al.}(2004)\citenamefont
  {Barbaro}, \citenamefont {Caballero}, \citenamefont {Donnelly},\ and\
  \citenamefont {Maieron}}]{PhysRevC.69.035502}%
  \BibitemOpen
  \bibfield  {author} {\bibinfo {author} {\bibfnamefont {M.~B.}\ \bibnamefont
  {Barbaro}}, \bibinfo {author} {\bibfnamefont {J.~A.}\ \bibnamefont
  {Caballero}}, \bibinfo {author} {\bibfnamefont {T.~W.}\ \bibnamefont
  {Donnelly}}, \ and\ \bibinfo {author} {\bibfnamefont {C.}~\bibnamefont
  {Maieron}},\ }\href {\doibase 10.1103/PhysRevC.69.035502} {\bibfield
  {journal} {\bibinfo  {journal} {Phys. Rev. C}\ }\textbf {\bibinfo {volume}
  {69}},\ \bibinfo {pages} {035502} (\bibinfo {year} {2004})}\BibitemShut
  {NoStop}%
\bibitem [{\citenamefont {Antonov}\ \emph
  {et~al.}(2006{\natexlab{a}})\citenamefont {Antonov}, \citenamefont {Ivanov},
  \citenamefont {Gaidarov}, \citenamefont {Guerra}, \citenamefont {Caballero},
  \citenamefont {Barbaro}, \citenamefont {Udias},\ and\ \citenamefont
  {Sarriguren}}]{PhysRevC.74.054603}%
  \BibitemOpen
  \bibfield  {author} {\bibinfo {author} {\bibfnamefont {A.~N.}\ \bibnamefont
  {Antonov}}, \bibinfo {author} {\bibfnamefont {M.~V.}\ \bibnamefont {Ivanov}},
  \bibinfo {author} {\bibfnamefont {M.~K.}\ \bibnamefont {Gaidarov}}, \bibinfo
  {author} {\bibfnamefont {E.~M.~d.}\ \bibnamefont {Guerra}}, \bibinfo {author}
  {\bibfnamefont {J.~A.}\ \bibnamefont {Caballero}}, \bibinfo {author}
  {\bibfnamefont {M.~B.}\ \bibnamefont {Barbaro}}, \bibinfo {author}
  {\bibfnamefont {J.~M.}\ \bibnamefont {Udias}}, \ and\ \bibinfo {author}
  {\bibfnamefont {P.}~\bibnamefont {Sarriguren}},\ }\href {\doibase
  10.1103/PhysRevC.74.054603} {\bibfield  {journal} {\bibinfo  {journal} {Phys.
  Rev. C}\ }\textbf {\bibinfo {volume} {74}},\ \bibinfo {pages} {054603}
  (\bibinfo {year} {2006}{\natexlab{a}})}\BibitemShut {NoStop}%
\bibitem [{\citenamefont {Antonov}\ \emph {et~al.}(2004)\citenamefont
  {Antonov}, \citenamefont {Gaidarov}, \citenamefont {Kadrev}, \citenamefont
  {Ivanov}, \citenamefont {Moya~de Guerra},\ and\ \citenamefont
  {Udias}}]{PhysRevC.69.044321}%
  \BibitemOpen
  \bibfield  {author} {\bibinfo {author} {\bibfnamefont {A.~N.}\ \bibnamefont
  {Antonov}}, \bibinfo {author} {\bibfnamefont {M.~K.}\ \bibnamefont
  {Gaidarov}}, \bibinfo {author} {\bibfnamefont {D.~N.}\ \bibnamefont
  {Kadrev}}, \bibinfo {author} {\bibfnamefont {M.~V.}\ \bibnamefont {Ivanov}},
  \bibinfo {author} {\bibfnamefont {E.}~\bibnamefont {Moya~de Guerra}}, \ and\
  \bibinfo {author} {\bibfnamefont {J.~M.}\ \bibnamefont {Udias}},\ }\href
  {\doibase 10.1103/PhysRevC.69.044321} {\bibfield  {journal} {\bibinfo
  {journal} {Phys. Rev. C}\ }\textbf {\bibinfo {volume} {69}},\ \bibinfo
  {pages} {044321} (\bibinfo {year} {2004})}\BibitemShut {NoStop}%
\bibitem [{\citenamefont {Antonov}\ \emph {et~al.}(2005)\citenamefont
  {Antonov}, \citenamefont {Gaidarov}, \citenamefont {Ivanov}, \citenamefont
  {Kadrev}, \citenamefont {Moya~de Guerra}, \citenamefont {Sarriguren},\ and\
  \citenamefont {Udias}}]{PhysRevC.71.014317}%
  \BibitemOpen
  \bibfield  {author} {\bibinfo {author} {\bibfnamefont {A.~N.}\ \bibnamefont
  {Antonov}}, \bibinfo {author} {\bibfnamefont {M.~K.}\ \bibnamefont
  {Gaidarov}}, \bibinfo {author} {\bibfnamefont {M.~V.}\ \bibnamefont
  {Ivanov}}, \bibinfo {author} {\bibfnamefont {D.~N.}\ \bibnamefont {Kadrev}},
  \bibinfo {author} {\bibfnamefont {E.}~\bibnamefont {Moya~de Guerra}},
  \bibinfo {author} {\bibfnamefont {P.}~\bibnamefont {Sarriguren}}, \ and\
  \bibinfo {author} {\bibfnamefont {J.~M.}\ \bibnamefont {Udias}},\ }\href
  {\doibase 10.1103/PhysRevC.71.014317} {\bibfield  {journal} {\bibinfo
  {journal} {Phys. Rev. C}\ }\textbf {\bibinfo {volume} {71}},\ \bibinfo
  {pages} {014317} (\bibinfo {year} {2005})}\BibitemShut {NoStop}%
\bibitem [{\citenamefont {Antonov}\ \emph
  {et~al.}(2006{\natexlab{b}})\citenamefont {Antonov}, \citenamefont {Ivanov},
  \citenamefont {Gaidarov}, \citenamefont {Moya~de Guerra}, \citenamefont
  {Sarriguren},\ and\ \citenamefont {Udias}}]{PhysRevC.73.047302}%
  \BibitemOpen
  \bibfield  {author} {\bibinfo {author} {\bibfnamefont {A.~N.}\ \bibnamefont
  {Antonov}}, \bibinfo {author} {\bibfnamefont {M.~V.}\ \bibnamefont {Ivanov}},
  \bibinfo {author} {\bibfnamefont {M.~K.}\ \bibnamefont {Gaidarov}}, \bibinfo
  {author} {\bibfnamefont {E.}~\bibnamefont {Moya~de Guerra}}, \bibinfo
  {author} {\bibfnamefont {P.}~\bibnamefont {Sarriguren}}, \ and\ \bibinfo
  {author} {\bibfnamefont {J.~M.}\ \bibnamefont {Udias}},\ }\href {\doibase
  10.1103/PhysRevC.73.047302} {\bibfield  {journal} {\bibinfo  {journal} {Phys.
  Rev. C}\ }\textbf {\bibinfo {volume} {73}},\ \bibinfo {pages} {047302}
  (\bibinfo {year} {2006}{\natexlab{b}})}\BibitemShut {NoStop}%
\bibitem [{\citenamefont {Ivanov}\ \emph {et~al.}(2008)\citenamefont {Ivanov},
  \citenamefont {Barbaro}, \citenamefont {Caballero}, \citenamefont {Antonov},
  \citenamefont {Moya~de Guerra},\ and\ \citenamefont
  {Gaidarov}}]{PhysRevC.77.034612}%
  \BibitemOpen
  \bibfield  {author} {\bibinfo {author} {\bibfnamefont {M.~V.}\ \bibnamefont
  {Ivanov}}, \bibinfo {author} {\bibfnamefont {M.~B.}\ \bibnamefont {Barbaro}},
  \bibinfo {author} {\bibfnamefont {J.~A.}\ \bibnamefont {Caballero}}, \bibinfo
  {author} {\bibfnamefont {A.~N.}\ \bibnamefont {Antonov}}, \bibinfo {author}
  {\bibfnamefont {E.}~\bibnamefont {Moya~de Guerra}}, \ and\ \bibinfo {author}
  {\bibfnamefont {M.~K.}\ \bibnamefont {Gaidarov}},\ }\href {\doibase
  10.1103/PhysRevC.77.034612} {\bibfield  {journal} {\bibinfo  {journal} {Phys.
  Rev. C}\ }\textbf {\bibinfo {volume} {77}},\ \bibinfo {pages} {034612}
  (\bibinfo {year} {2008})}\BibitemShut {NoStop}%
\bibitem [{\citenamefont {Ruiz~Simo}\ \emph {et~al.}(2018)\citenamefont
  {Ruiz~Simo}, \citenamefont {Martinez-Consentino}, \citenamefont {Amaro},\
  and\ \citenamefont {Ruiz~Arriola}}]{RuizSimo:2018kdl}%
  \BibitemOpen
  \bibfield  {author} {\bibinfo {author} {\bibfnamefont {I.}~\bibnamefont
  {Ruiz~Simo}}, \bibinfo {author} {\bibfnamefont {V.~L.}\ \bibnamefont
  {Martinez-Consentino}}, \bibinfo {author} {\bibfnamefont {J.~E.}\
  \bibnamefont {Amaro}}, \ and\ \bibinfo {author} {\bibfnamefont
  {E.}~\bibnamefont {Ruiz~Arriola}},\ }\href {\doibase
  10.1103/PhysRevD.97.116006} {\bibfield  {journal} {\bibinfo  {journal} {Phys.
  Rev. D}\ }\textbf {\bibinfo {volume} {97}},\ \bibinfo {pages} {116006}
  (\bibinfo {year} {2018})}\BibitemShut {NoStop}%
\bibitem [{\citenamefont {Martini}\ \emph {et~al.}(2009)\citenamefont
  {Martini}, \citenamefont {Ericson}, \citenamefont {Chanfray},\ and\
  \citenamefont {Marteau}}]{Martini:2009uj}%
  \BibitemOpen
  \bibfield  {author} {\bibinfo {author} {\bibfnamefont {M.}~\bibnamefont
  {Martini}}, \bibinfo {author} {\bibfnamefont {M.}~\bibnamefont {Ericson}},
  \bibinfo {author} {\bibfnamefont {G.}~\bibnamefont {Chanfray}}, \ and\
  \bibinfo {author} {\bibfnamefont {J.}~\bibnamefont {Marteau}},\ }\href
  {\doibase 10.1103/PhysRevC.80.065501} {\bibfield  {journal} {\bibinfo
  {journal} {Phys. Rev. C}\ }\textbf {\bibinfo {volume} {80}},\ \bibinfo
  {pages} {065501} (\bibinfo {year} {2009})}\BibitemShut {NoStop}%
\bibitem [{\citenamefont {Amaro}\ \emph {et~al.}(2011)\citenamefont {Amaro},
  \citenamefont {Barbaro}, \citenamefont {Caballero}, \citenamefont
  {Donnelly},\ and\ \citenamefont {Williamson}}]{Amaro:2010sd}%
  \BibitemOpen
  \bibfield  {author} {\bibinfo {author} {\bibfnamefont {J.~E.}\ \bibnamefont
  {Amaro}}, \bibinfo {author} {\bibfnamefont {M.~B.}\ \bibnamefont {Barbaro}},
  \bibinfo {author} {\bibfnamefont {J.~A.}\ \bibnamefont {Caballero}}, \bibinfo
  {author} {\bibfnamefont {T.~W.}\ \bibnamefont {Donnelly}}, \ and\ \bibinfo
  {author} {\bibfnamefont {C.~F.}\ \bibnamefont {Williamson}},\ }\href
  {\doibase 10.1016/j.physletb.2010.12.007} {\bibfield  {journal} {\bibinfo
  {journal} {Phys. Lett. B}\ }\textbf {\bibinfo {volume} {696}},\ \bibinfo
  {pages} {151} (\bibinfo {year} {2011})}\BibitemShut {NoStop}%
\bibitem [{\citenamefont {Nieves}\ \emph
  {et~al.}(2012{\natexlab{a}})\citenamefont {Nieves}, \citenamefont
  {Ruiz~Simo},\ and\ \citenamefont {Vicente~Vacas}}]{Nieves:2011yp}%
  \BibitemOpen
  \bibfield  {author} {\bibinfo {author} {\bibfnamefont {J.}~\bibnamefont
  {Nieves}}, \bibinfo {author} {\bibfnamefont {I.}~\bibnamefont {Ruiz~Simo}}, \
  and\ \bibinfo {author} {\bibfnamefont {M.~J.}\ \bibnamefont
  {Vicente~Vacas}},\ }\href {\doibase 10.1016/j.physletb.2011.11.061}
  {\bibfield  {journal} {\bibinfo  {journal} {Physics Letters B}\ }\textbf
  {\bibinfo {volume} {707}},\ \bibinfo {pages} {72 } (\bibinfo {year}
  {2012}{\natexlab{a}})}\BibitemShut {NoStop}%
\bibitem [{\citenamefont {Lalakulich}\ \emph {et~al.}(2012)\citenamefont
  {Lalakulich}, \citenamefont {Gallmeister},\ and\ \citenamefont
  {Mosel}}]{PhysRevC.86.014614}%
  \BibitemOpen
  \bibfield  {author} {\bibinfo {author} {\bibfnamefont {O.}~\bibnamefont
  {Lalakulich}}, \bibinfo {author} {\bibfnamefont {K.}~\bibnamefont
  {Gallmeister}}, \ and\ \bibinfo {author} {\bibfnamefont {U.}~\bibnamefont
  {Mosel}},\ }\href {\doibase 10.1103/PhysRevC.86.014614} {\bibfield  {journal}
  {\bibinfo  {journal} {Phys. Rev. C}\ }\textbf {\bibinfo {volume} {86}},\
  \bibinfo {pages} {014614} (\bibinfo {year} {2012})}\BibitemShut {NoStop}%
\bibitem [{\citenamefont {Gonz\'alez-Jim\'enez}\ \emph
  {et~al.}(2014)\citenamefont {Gonz\'alez-Jim\'enez}, \citenamefont {Megias},
  \citenamefont {Barbaro}, \citenamefont {Caballero},\ and\ \citenamefont
  {Donnelly}}]{PhysRevC.90.035501}%
  \BibitemOpen
  \bibfield  {author} {\bibinfo {author} {\bibfnamefont {R.}~\bibnamefont
  {Gonz\'alez-Jim\'enez}}, \bibinfo {author} {\bibfnamefont {G.~D.}\
  \bibnamefont {Megias}}, \bibinfo {author} {\bibfnamefont {M.~B.}\
  \bibnamefont {Barbaro}}, \bibinfo {author} {\bibfnamefont {J.~A.}\
  \bibnamefont {Caballero}}, \ and\ \bibinfo {author} {\bibfnamefont {T.~W.}\
  \bibnamefont {Donnelly}},\ }\href {\doibase 10.1103/PhysRevC.90.035501}
  {\bibfield  {journal} {\bibinfo  {journal} {Phys. Rev. C}\ }\textbf {\bibinfo
  {volume} {90}},\ \bibinfo {pages} {035501} (\bibinfo {year}
  {2014})}\BibitemShut {NoStop}%
\bibitem [{\citenamefont {Ivanov}\ \emph {et~al.}(2014)\citenamefont {Ivanov},
  \citenamefont {Antonov}, \citenamefont {Caballero}, \citenamefont {Megias},
  \citenamefont {Barbaro}, \citenamefont {de~Guerra},\ and\ \citenamefont
  {Ud\'{\i}as}}]{PhysRevC.89.014607}%
  \BibitemOpen
  \bibfield  {author} {\bibinfo {author} {\bibfnamefont {M.~V.}\ \bibnamefont
  {Ivanov}}, \bibinfo {author} {\bibfnamefont {A.~N.}\ \bibnamefont {Antonov}},
  \bibinfo {author} {\bibfnamefont {J.~A.}\ \bibnamefont {Caballero}}, \bibinfo
  {author} {\bibfnamefont {G.~D.}\ \bibnamefont {Megias}}, \bibinfo {author}
  {\bibfnamefont {M.~B.}\ \bibnamefont {Barbaro}}, \bibinfo {author}
  {\bibfnamefont {E.~M.}\ \bibnamefont {de~Guerra}}, \ and\ \bibinfo {author}
  {\bibfnamefont {J.~M.}\ \bibnamefont {Ud\'{\i}as}},\ }\href {\doibase
  10.1103/PhysRevC.89.014607} {\bibfield  {journal} {\bibinfo  {journal} {Phys.
  Rev. C}\ }\textbf {\bibinfo {volume} {89}},\ \bibinfo {pages} {014607}
  (\bibinfo {year} {2014})}\BibitemShut {NoStop}%
\bibitem [{\citenamefont {Ivanov}\ \emph {et~al.}(2015)\citenamefont {Ivanov},
  \citenamefont {Antonov}, \citenamefont {Barbaro}, \citenamefont {Giusti},
  \citenamefont {Meucci}, \citenamefont {Caballero}, \citenamefont
  {Gonz\'alez-Jim\'enez}, \citenamefont {de~Guerra},\ and\ \citenamefont
  {Ud\'{\i}as}}]{PhysRevC.91.034607}%
  \BibitemOpen
  \bibfield  {author} {\bibinfo {author} {\bibfnamefont {M.~V.}\ \bibnamefont
  {Ivanov}}, \bibinfo {author} {\bibfnamefont {A.~N.}\ \bibnamefont {Antonov}},
  \bibinfo {author} {\bibfnamefont {M.~B.}\ \bibnamefont {Barbaro}}, \bibinfo
  {author} {\bibfnamefont {C.}~\bibnamefont {Giusti}}, \bibinfo {author}
  {\bibfnamefont {A.}~\bibnamefont {Meucci}}, \bibinfo {author} {\bibfnamefont
  {J.~A.}\ \bibnamefont {Caballero}}, \bibinfo {author} {\bibfnamefont
  {R.}~\bibnamefont {Gonz\'alez-Jim\'enez}}, \bibinfo {author} {\bibfnamefont
  {E.~M.}\ \bibnamefont {de~Guerra}}, \ and\ \bibinfo {author} {\bibfnamefont
  {J.~M.}\ \bibnamefont {Ud\'{\i}as}},\ }\href {\doibase
  10.1103/PhysRevC.91.034607} {\bibfield  {journal} {\bibinfo  {journal} {Phys.
  Rev. C}\ }\textbf {\bibinfo {volume} {91}},\ \bibinfo {pages} {034607}
  (\bibinfo {year} {2015})}\BibitemShut {NoStop}%
\bibitem [{\citenamefont {Antonov}\ \emph {et~al.}(2011)\citenamefont
  {Antonov}, \citenamefont {Ivanov}, \citenamefont {Caballero}, \citenamefont
  {Barbaro}, \citenamefont {Udias}, \citenamefont {Moya~de Guerra},\ and\
  \citenamefont {Donnelly}}]{PhysRevC.83.045504}%
  \BibitemOpen
  \bibfield  {author} {\bibinfo {author} {\bibfnamefont {A.~N.}\ \bibnamefont
  {Antonov}}, \bibinfo {author} {\bibfnamefont {M.~V.}\ \bibnamefont {Ivanov}},
  \bibinfo {author} {\bibfnamefont {J.~A.}\ \bibnamefont {Caballero}}, \bibinfo
  {author} {\bibfnamefont {M.~B.}\ \bibnamefont {Barbaro}}, \bibinfo {author}
  {\bibfnamefont {J.~M.}\ \bibnamefont {Udias}}, \bibinfo {author}
  {\bibfnamefont {E.}~\bibnamefont {Moya~de Guerra}}, \ and\ \bibinfo {author}
  {\bibfnamefont {T.~W.}\ \bibnamefont {Donnelly}},\ }\href {\doibase
  10.1103/PhysRevC.83.045504} {\bibfield  {journal} {\bibinfo  {journal} {Phys.
  Rev. C}\ }\textbf {\bibinfo {volume} {83}},\ \bibinfo {pages} {045504}
  (\bibinfo {year} {2011})}\BibitemShut {NoStop}%
\bibitem [{\citenamefont {Simo}\ \emph {et~al.}(2014)\citenamefont {Simo},
  \citenamefont {Albertus}, \citenamefont {Amaro}, \citenamefont {Barbaro},
  \citenamefont {Caballero},\ and\ \citenamefont
  {Donnelly}}]{PhysRevD.90.033012}%
  \BibitemOpen
  \bibfield  {author} {\bibinfo {author} {\bibfnamefont {I.~R.}\ \bibnamefont
  {Simo}}, \bibinfo {author} {\bibfnamefont {C.}~\bibnamefont {Albertus}},
  \bibinfo {author} {\bibfnamefont {J.~E.}\ \bibnamefont {Amaro}}, \bibinfo
  {author} {\bibfnamefont {M.~B.}\ \bibnamefont {Barbaro}}, \bibinfo {author}
  {\bibfnamefont {J.~A.}\ \bibnamefont {Caballero}}, \ and\ \bibinfo {author}
  {\bibfnamefont {T.~W.}\ \bibnamefont {Donnelly}},\ }\href {\doibase
  10.1103/PhysRevD.90.033012} {\bibfield  {journal} {\bibinfo  {journal} {Phys.
  Rev. D}\ }\textbf {\bibinfo {volume} {90}},\ \bibinfo {pages} {033012}
  (\bibinfo {year} {2014})}\BibitemShut {NoStop}%
\bibitem [{\citenamefont {Ruiz~Simo}\ \emph {et~al.}(2017)\citenamefont
  {Ruiz~Simo}, \citenamefont {Amaro}, \citenamefont {Barbaro}, \citenamefont
  {De~Pace}, \citenamefont {Caballero},\ and\ \citenamefont
  {Donnelly}}]{Simo:2016ikv}%
  \BibitemOpen
  \bibfield  {author} {\bibinfo {author} {\bibfnamefont {I.}~\bibnamefont
  {Ruiz~Simo}}, \bibinfo {author} {\bibfnamefont {J.~E.}\ \bibnamefont
  {Amaro}}, \bibinfo {author} {\bibfnamefont {M.~B.}\ \bibnamefont {Barbaro}},
  \bibinfo {author} {\bibfnamefont {A.}~\bibnamefont {De~Pace}}, \bibinfo
  {author} {\bibfnamefont {J.~A.}\ \bibnamefont {Caballero}}, \ and\ \bibinfo
  {author} {\bibfnamefont {T.~W.}\ \bibnamefont {Donnelly}},\ }\href {\doibase
  10.1088/1361-6471/aa6a06} {\bibfield  {journal} {\bibinfo  {journal} {J.
  Phys.}\ }\textbf {\bibinfo {volume} {G44}},\ \bibinfo {pages} {065105}
  (\bibinfo {year} {2017})},\ \Eprint {http://arxiv.org/abs/1604.08423}
  {arXiv:1604.08423 [nucl-th]} \BibitemShut {NoStop}%
%%CITATION = ARXIV:1604.08423;%%
\bibitem [{\citenamefont {Simo}\ \emph {et~al.}(2016)\citenamefont {Simo},
  \citenamefont {Amaro}, \citenamefont {Barbaro}, \citenamefont {De~Pace},
  \citenamefont {Caballero}, \citenamefont {Megias},\ and\ \citenamefont
  {Donnelly}}]{PhysRevC.94.054610}%
  \BibitemOpen
  \bibfield  {author} {\bibinfo {author} {\bibfnamefont {I.~R.}\ \bibnamefont
  {Simo}}, \bibinfo {author} {\bibfnamefont {J.~E.}\ \bibnamefont {Amaro}},
  \bibinfo {author} {\bibfnamefont {M.~B.}\ \bibnamefont {Barbaro}}, \bibinfo
  {author} {\bibfnamefont {A.}~\bibnamefont {De~Pace}}, \bibinfo {author}
  {\bibfnamefont {J.~A.}\ \bibnamefont {Caballero}}, \bibinfo {author}
  {\bibfnamefont {G.~D.}\ \bibnamefont {Megias}}, \ and\ \bibinfo {author}
  {\bibfnamefont {T.~W.}\ \bibnamefont {Donnelly}},\ }\href {\doibase
  10.1103/PhysRevC.94.054610} {\bibfield  {journal} {\bibinfo  {journal} {Phys.
  Rev. C}\ }\textbf {\bibinfo {volume} {94}},\ \bibinfo {pages} {054610}
  (\bibinfo {year} {2016})}\BibitemShut {NoStop}%
\bibitem [{\citenamefont {Aguilar-Arevalo}\ \emph {et~al.}(2010)\citenamefont
  {Aguilar-Arevalo} \emph {et~al.}}]{miniboone}%
  \BibitemOpen
  \bibfield  {author} {\bibinfo {author} {\bibfnamefont {A.~A.}\ \bibnamefont
  {Aguilar-Arevalo}} \emph {et~al.} (\bibinfo {collaboration} {MiniBooNE
  Collaboration}),\ }\href@noop {} {\bibfield  {journal} {\bibinfo  {journal}
  {Phys. Rev. D}\ }\textbf {\bibinfo {volume} {81}},\ \bibinfo {pages} {092005}
  (\bibinfo {year} {2010})}\BibitemShut {NoStop}%
\bibitem [{\citenamefont {Aguilar-Arevalo}\ \emph {et~al.}(2013)\citenamefont
  {Aguilar-Arevalo} \emph {et~al.}}]{miniboone-ant}%
  \BibitemOpen
  \bibfield  {author} {\bibinfo {author} {\bibfnamefont {A.~A.}\ \bibnamefont
  {Aguilar-Arevalo}} \emph {et~al.} (\bibinfo {collaboration} {MiniBooNE
  Collaboration}),\ }\href@noop {} {\bibfield  {journal} {\bibinfo  {journal}
  {Phys. Rev. D}\ }\textbf {\bibinfo {volume} {88}},\ \bibinfo {pages} {032001}
  (\bibinfo {year} {2013})}\BibitemShut {NoStop}%
\bibitem [{\citenamefont {Lyubushkin}\ \emph {et~al.}(2009)\citenamefont
  {Lyubushkin} \emph {et~al.}}]{Lyubushkin:2009}%
  \BibitemOpen
  \bibfield  {author} {\bibinfo {author} {\bibfnamefont {V.}~\bibnamefont
  {Lyubushkin}} \emph {et~al.} (\bibinfo {collaboration} {NOMAD
  Collaboration}),\ }\href {\doibase 10.1140/epjc/s10052-009-1113-0} {\bibfield
   {journal} {\bibinfo  {journal} {Eur. Phys. J. C}\ }\textbf {\bibinfo
  {volume} {63}},\ \bibinfo {pages} {355} (\bibinfo {year} {2009})}\BibitemShut
  {NoStop}%
\bibitem [{\citenamefont {Abe}\ \emph {et~al.}(2016)\citenamefont {Abe} \emph
  {et~al.}}]{PhysRevD.93.112012}%
  \BibitemOpen
  \bibfield  {author} {\bibinfo {author} {\bibfnamefont {K.}~\bibnamefont
  {Abe}} \emph {et~al.} (\bibinfo {collaboration} {T2K Collaboration}),\ }\href
  {\doibase 10.1103/PhysRevD.93.112012} {\bibfield  {journal} {\bibinfo
  {journal} {Phys. Rev. D}\ }\textbf {\bibinfo {volume} {93}},\ \bibinfo
  {pages} {112012} (\bibinfo {year} {2016})}\BibitemShut {NoStop}%
\bibitem [{\citenamefont {Fiorentini}\ \emph {et~al.}(2013)\citenamefont
  {Fiorentini} \emph {et~al.}}]{PhysRevLett.111.022502}%
  \BibitemOpen
  \bibfield  {author} {\bibinfo {author} {\bibfnamefont {G.~A.}\ \bibnamefont
  {Fiorentini}} \emph {et~al.} (\bibinfo {collaboration} {MINERvA
  Collaboration}),\ }\href {\doibase 10.1103/PhysRevLett.111.022502} {\bibfield
   {journal} {\bibinfo  {journal} {Phys. Rev. Lett.}\ }\textbf {\bibinfo
  {volume} {111}},\ \bibinfo {pages} {022502} (\bibinfo {year} {2013})},\
  \bibinfo {note} {the values come from a private communication, MINERvA
  collaboration.}\BibitemShut {Stop}%
\bibitem [{\citenamefont {Fields}\ \emph {et~al.}(2013)\citenamefont {Fields}
  \emph {et~al.}}]{PhysRevLett.111.022501}%
  \BibitemOpen
  \bibfield  {author} {\bibinfo {author} {\bibfnamefont {L.}~\bibnamefont
  {Fields}} \emph {et~al.} (\bibinfo {collaboration} {MINERvA Collaboration}),\
  }\href {\doibase 10.1103/PhysRevLett.111.022501} {\bibfield  {journal}
  {\bibinfo  {journal} {Phys. Rev. Lett.}\ }\textbf {\bibinfo {volume} {111}},\
  \bibinfo {pages} {022501} (\bibinfo {year} {2013})},\ \bibinfo {note} {the
  values come from a private communication, MINERvA collaboration.}\BibitemShut
  {Stop}%
\bibitem [{\citenamefont {Patrick}(2016)}]{cherylthesis}%
  \BibitemOpen
  \bibfield  {author} {\bibinfo {author} {\bibfnamefont {C.}~\bibnamefont
  {Patrick}},\ }\emph {\bibinfo {title} {Measurement of the Antineutrino
  Double-Differential Charged-Current Quasi-Elastic Scattering Cross Section at
  MINER$\nu$A}},\ \href
  {http://lss.fnal.gov/archive/thesis/2000/fermilab-thesis-2016-04.pdf} {Ph.D.
  thesis},\ \bibinfo  {school} {Northwestern U.} (\bibinfo {year} {2016}),\
  \bibinfo {note} {available as FERMILAB-THESIS-2016-04}\BibitemShut {NoStop}%
\bibitem [{\citenamefont {Patrick}\ \emph {et~al.}(2018)\citenamefont {Patrick}
  \emph {et~al.}}]{PhysRevD.97.052002}%
  \BibitemOpen
  \bibfield  {author} {\bibinfo {author} {\bibfnamefont {C.~E.}\ \bibnamefont
  {Patrick}} \emph {et~al.} (\bibinfo {collaboration} {MINER\ensuremath{\nu}A
  Collaboration}),\ }\href {\doibase 10.1103/PhysRevD.97.052002} {\bibfield
  {journal} {\bibinfo  {journal} {Phys. Rev. D}\ }\textbf {\bibinfo {volume}
  {97}},\ \bibinfo {pages} {052002} (\bibinfo {year} {2018})}\BibitemShut
  {NoStop}%
\bibitem [{\citenamefont {L\"owdin}(1955)}]{Lowdin:1955}%
  \BibitemOpen
  \bibfield  {author} {\bibinfo {author} {\bibfnamefont {P.-O.}\ \bibnamefont
  {L\"owdin}},\ }\href {\doibase 10.1103/PhysRev.97.1474} {\bibfield  {journal}
  {\bibinfo  {journal} {Phys. Rev.}\ }\textbf {\bibinfo {volume} {97}},\
  \bibinfo {pages} {1474} (\bibinfo {year} {1955})}\BibitemShut {NoStop}%
\bibitem [{\citenamefont {Stoitsov}\ \emph {et~al.}(1993)\citenamefont
  {Stoitsov}, \citenamefont {Antonov},\ and\ \citenamefont
  {Dimitrova}}]{Stoitsov:1993}%
  \BibitemOpen
  \bibfield  {author} {\bibinfo {author} {\bibfnamefont {M.~V.}\ \bibnamefont
  {Stoitsov}}, \bibinfo {author} {\bibfnamefont {A.~N.}\ \bibnamefont
  {Antonov}}, \ and\ \bibinfo {author} {\bibfnamefont {S.~S.}\ \bibnamefont
  {Dimitrova}},\ }\href {\doibase 10.1103/PhysRevC.48.74} {\bibfield  {journal}
  {\bibinfo  {journal} {Phys. Rev. C}\ }\textbf {\bibinfo {volume} {48}},\
  \bibinfo {pages} {74} (\bibinfo {year} {1993})}\BibitemShut {NoStop}%
\bibitem [{\citenamefont {Dutta}(1999)}]{Dutta:1999}%
  \BibitemOpen
  \bibfield  {author} {\bibinfo {author} {\bibfnamefont {D.}~\bibnamefont
  {Dutta}},\ }\emph {\bibinfo {title} {The ($e,e'p$) Reaction Mechanism in the
  Quasi-Elastic Region}},\ \href@noop {} {Ph.D. thesis},\ \bibinfo  {school}
  {Northwestern University} (\bibinfo {year} {1999})\BibitemShut {NoStop}%
\bibitem [{\citenamefont {Ankowski}\ and\ \citenamefont
  {Sobczyk}(2008)}]{PhysRevC.77.044311}%
  \BibitemOpen
  \bibfield  {author} {\bibinfo {author} {\bibfnamefont {A.~M.}\ \bibnamefont
  {Ankowski}}\ and\ \bibinfo {author} {\bibfnamefont {J.~T.}\ \bibnamefont
  {Sobczyk}},\ }\href {\doibase 10.1103/PhysRevC.77.044311} {\bibfield
  {journal} {\bibinfo  {journal} {Phys. Rev. C}\ }\textbf {\bibinfo {volume}
  {77}},\ \bibinfo {pages} {044311} (\bibinfo {year} {2008})}\BibitemShut
  {NoStop}%
\bibitem [{\citenamefont {Horikawa}\ \emph {et~al.}(1980)\citenamefont
  {Horikawa}, \citenamefont {Lenz},\ and\ \citenamefont
  {Mukhopadhyay}}]{PhysRevC.22.1680}%
  \BibitemOpen
  \bibfield  {author} {\bibinfo {author} {\bibfnamefont {Y.}~\bibnamefont
  {Horikawa}}, \bibinfo {author} {\bibfnamefont {F.}~\bibnamefont {Lenz}}, \
  and\ \bibinfo {author} {\bibfnamefont {N.~C.}\ \bibnamefont {Mukhopadhyay}},\
  }\href {\doibase 10.1103/PhysRevC.22.1680} {\bibfield  {journal} {\bibinfo
  {journal} {Phys. Rev. C}\ }\textbf {\bibinfo {volume} {22}},\ \bibinfo
  {pages} {1680} (\bibinfo {year} {1980})}\BibitemShut {NoStop}%
\bibitem [{\citenamefont {Clark}\ \emph {et~al.}(2006)\citenamefont {Clark},
  \citenamefont {Cooper},\ and\ \citenamefont {Hama}}]{PhysRevC.73.024608}%
  \BibitemOpen
  \bibfield  {author} {\bibinfo {author} {\bibfnamefont {B.~C.}\ \bibnamefont
  {Clark}}, \bibinfo {author} {\bibfnamefont {E.~D.}\ \bibnamefont {Cooper}}, \
  and\ \bibinfo {author} {\bibfnamefont {S.}~\bibnamefont {Hama}},\ }\href
  {\doibase 10.1103/PhysRevC.73.024608} {\bibfield  {journal} {\bibinfo
  {journal} {Phys. Rev. C}\ }\textbf {\bibinfo {volume} {73}},\ \bibinfo
  {pages} {024608} (\bibinfo {year} {2006})}\BibitemShut {NoStop}%
\bibitem [{\citenamefont {Rosenfelder}(1980)}]{ROSENFELDER1980188}%
  \BibitemOpen
  \bibfield  {author} {\bibinfo {author} {\bibfnamefont {R.}~\bibnamefont
  {Rosenfelder}},\ }\href {\doibase
  https://doi.org/10.1016/0003-4916(80)90059-7} {\bibfield  {journal} {\bibinfo
   {journal} {Annals of Physics}\ }\textbf {\bibinfo {volume} {128}},\ \bibinfo
  {pages} {188 } (\bibinfo {year} {1980})}\BibitemShut {NoStop}%
\bibitem [{\citenamefont {Megias}\ \emph {et~al.}(2014)\citenamefont {Megias},
  \citenamefont {Ivanov}, \citenamefont {Gonz\'alez-Jim\'enez}, \citenamefont
  {Barbaro}, \citenamefont {Caballero}, \citenamefont {Donnelly},\ and\
  \citenamefont {Ud\'{\i}as}}]{PhysRevD.89.093002}%
  \BibitemOpen
  \bibfield  {author} {\bibinfo {author} {\bibfnamefont {G.~D.}\ \bibnamefont
  {Megias}}, \bibinfo {author} {\bibfnamefont {M.~V.}\ \bibnamefont {Ivanov}},
  \bibinfo {author} {\bibfnamefont {R.}~\bibnamefont {Gonz\'alez-Jim\'enez}},
  \bibinfo {author} {\bibfnamefont {M.~B.}\ \bibnamefont {Barbaro}}, \bibinfo
  {author} {\bibfnamefont {J.~A.}\ \bibnamefont {Caballero}}, \bibinfo {author}
  {\bibfnamefont {T.~W.}\ \bibnamefont {Donnelly}}, \ and\ \bibinfo {author}
  {\bibfnamefont {J.~M.}\ \bibnamefont {Ud\'{\i}as}},\ }\href {\doibase
  10.1103/PhysRevD.89.093002} {\bibfield  {journal} {\bibinfo  {journal} {Phys.
  Rev. D}\ }\textbf {\bibinfo {volume} {89}},\ \bibinfo {pages} {093002}
  (\bibinfo {year} {2014})}\BibitemShut {NoStop}%
\bibitem [{\citenamefont {Sobczyk}\ \emph {et~al.}(2018)\citenamefont
  {Sobczyk}, \citenamefont {Rocco}, \citenamefont {Lovato},\ and\ \citenamefont
  {Nieves}}]{PhysRevC.97.035506}%
  \BibitemOpen
  \bibfield  {author} {\bibinfo {author} {\bibfnamefont {J.~E.}\ \bibnamefont
  {Sobczyk}}, \bibinfo {author} {\bibfnamefont {N.}~\bibnamefont {Rocco}},
  \bibinfo {author} {\bibfnamefont {A.}~\bibnamefont {Lovato}}, \ and\ \bibinfo
  {author} {\bibfnamefont {J.}~\bibnamefont {Nieves}},\ }\href {\doibase
  10.1103/PhysRevC.97.035506} {\bibfield  {journal} {\bibinfo  {journal} {Phys.
  Rev. C}\ }\textbf {\bibinfo {volume} {97}},\ \bibinfo {pages} {035506}
  (\bibinfo {year} {2018})}\BibitemShut {NoStop}%
\bibitem [{\citenamefont {Caballero}\ \emph {et~al.}(2005)\citenamefont
  {Caballero}, \citenamefont {Amaro}, \citenamefont {Barbaro}, \citenamefont
  {Donnelly}, \citenamefont {Maieron},\ and\ \citenamefont
  {Udias}}]{PhysRevLett.95.252502}%
  \BibitemOpen
  \bibfield  {author} {\bibinfo {author} {\bibfnamefont {J.~A.}\ \bibnamefont
  {Caballero}}, \bibinfo {author} {\bibfnamefont {J.~E.}\ \bibnamefont
  {Amaro}}, \bibinfo {author} {\bibfnamefont {M.~B.}\ \bibnamefont {Barbaro}},
  \bibinfo {author} {\bibfnamefont {T.~W.}\ \bibnamefont {Donnelly}}, \bibinfo
  {author} {\bibfnamefont {C.}~\bibnamefont {Maieron}}, \ and\ \bibinfo
  {author} {\bibfnamefont {J.~M.}\ \bibnamefont {Udias}},\ }\href {\doibase
  10.1103/PhysRevLett.95.252502} {\bibfield  {journal} {\bibinfo  {journal}
  {Phys. Rev. Lett.}\ }\textbf {\bibinfo {volume} {95}},\ \bibinfo {pages}
  {252502} (\bibinfo {year} {2005})}\BibitemShut {NoStop}%
\bibitem [{\citenamefont {Caballero}(2006)}]{PhysRevC.74.015502}%
  \BibitemOpen
  \bibfield  {author} {\bibinfo {author} {\bibfnamefont {J.~A.}\ \bibnamefont
  {Caballero}},\ }\href {\doibase 10.1103/PhysRevC.74.015502} {\bibfield
  {journal} {\bibinfo  {journal} {Phys. Rev. C}\ }\textbf {\bibinfo {volume}
  {74}},\ \bibinfo {pages} {015502} (\bibinfo {year} {2006})}\BibitemShut
  {NoStop}%
\bibitem [{\citenamefont {Caballero}\ \emph {et~al.}(2007)\citenamefont
  {Caballero}, \citenamefont {Amaro}, \citenamefont {Barbaro}, \citenamefont
  {Donnelly},\ and\ \citenamefont {Ud\'{\i}as}}]{CABALLERO2007366}%
  \BibitemOpen
  \bibfield  {author} {\bibinfo {author} {\bibfnamefont {J.}~\bibnamefont
  {Caballero}}, \bibinfo {author} {\bibfnamefont {J.}~\bibnamefont {Amaro}},
  \bibinfo {author} {\bibfnamefont {M.}~\bibnamefont {Barbaro}}, \bibinfo
  {author} {\bibfnamefont {T.}~\bibnamefont {Donnelly}}, \ and\ \bibinfo
  {author} {\bibfnamefont {J.}~\bibnamefont {Ud\'{\i}as}},\ }\href {\doibase
  https://doi.org/10.1016/j.physletb.2007.08.018} {\bibfield  {journal}
  {\bibinfo  {journal} {Physics Letters B}\ }\textbf {\bibinfo {volume}
  {653}},\ \bibinfo {pages} {366 } (\bibinfo {year} {2007})}\BibitemShut
  {NoStop}%
\bibitem [{\citenamefont {Megias}\ \emph
  {et~al.}(2016{\natexlab{b}})\citenamefont {Megias}, \citenamefont {Amaro},
  \citenamefont {Barbaro}, \citenamefont {Caballero},\ and\ \citenamefont
  {Donnelly}}]{PhysRevD.94.013012}%
  \BibitemOpen
  \bibfield  {author} {\bibinfo {author} {\bibfnamefont {G.~D.}\ \bibnamefont
  {Megias}}, \bibinfo {author} {\bibfnamefont {J.~E.}\ \bibnamefont {Amaro}},
  \bibinfo {author} {\bibfnamefont {M.~B.}\ \bibnamefont {Barbaro}}, \bibinfo
  {author} {\bibfnamefont {J.~A.}\ \bibnamefont {Caballero}}, \ and\ \bibinfo
  {author} {\bibfnamefont {T.~W.}\ \bibnamefont {Donnelly}},\ }\href {\doibase
  10.1103/PhysRevD.94.013012} {\bibfield  {journal} {\bibinfo  {journal} {Phys.
  Rev. D}\ }\textbf {\bibinfo {volume} {94}},\ \bibinfo {pages} {013012}
  (\bibinfo {year} {2016}{\natexlab{b}})}\BibitemShut {NoStop}%
\bibitem [{\citenamefont {Katori}\ and\ \citenamefont
  {Martini}(2018)}]{Katori:2016yel}%
  \BibitemOpen
  \bibfield  {author} {\bibinfo {author} {\bibfnamefont {T.}~\bibnamefont
  {Katori}}\ and\ \bibinfo {author} {\bibfnamefont {M.}~\bibnamefont
  {Martini}},\ }\href {\doibase 10.1088/1361-6471/aa8bf7} {\bibfield  {journal}
  {\bibinfo  {journal} {Journal of Physics G: Nuclear and Particle Physics}\
  }\textbf {\bibinfo {volume} {45}},\ \bibinfo {pages} {013001} (\bibinfo
  {year} {2018})}\BibitemShut {NoStop}%
\bibitem [{\citenamefont {Van~Orden}\ and\ \citenamefont
  {Donnelly}(1981)}]{VanOrden:1980tg}%
  \BibitemOpen
  \bibfield  {author} {\bibinfo {author} {\bibfnamefont {J.~W.}\ \bibnamefont
  {Van~Orden}}\ and\ \bibinfo {author} {\bibfnamefont {T.~W.}\ \bibnamefont
  {Donnelly}},\ }\href {\doibase 10.1016/0003-4916(81)90038-5} {\bibfield
  {journal} {\bibinfo  {journal} {Annals Phys.}\ }\textbf {\bibinfo {volume}
  {131}},\ \bibinfo {pages} {451} (\bibinfo {year} {1981})}\BibitemShut
  {NoStop}%
%%CITATION = APNYA,131,451;%%
\bibitem [{\citenamefont {De~Pace}\ \emph {et~al.}(2003)\citenamefont
  {De~Pace}, \citenamefont {Nardi}, \citenamefont {Alberico}, \citenamefont
  {Donnelly},\ and\ \citenamefont {Molinari}}]{DePace:2003spn}%
  \BibitemOpen
  \bibfield  {author} {\bibinfo {author} {\bibfnamefont {A.}~\bibnamefont
  {De~Pace}}, \bibinfo {author} {\bibfnamefont {M.}~\bibnamefont {Nardi}},
  \bibinfo {author} {\bibfnamefont {W.~M.}\ \bibnamefont {Alberico}}, \bibinfo
  {author} {\bibfnamefont {T.~W.}\ \bibnamefont {Donnelly}}, \ and\ \bibinfo
  {author} {\bibfnamefont {A.}~\bibnamefont {Molinari}},\ }\href {\doibase
  10.1016/S0375-9474(03)01625-7} {\bibfield  {journal} {\bibinfo  {journal}
  {Nucl. Phys.}\ }\textbf {\bibinfo {volume} {A726}},\ \bibinfo {pages} {303}
  (\bibinfo {year} {2003})}\BibitemShut {NoStop}%
\bibitem [{\citenamefont {Amaro}\ \emph {et~al.}(2010)\citenamefont {Amaro},
  \citenamefont {Maieron}, \citenamefont {Barbaro}, \citenamefont {Caballero},\
  and\ \citenamefont {Donnelly}}]{Amaro:2010iu}%
  \BibitemOpen
  \bibfield  {author} {\bibinfo {author} {\bibfnamefont {J.~E.}\ \bibnamefont
  {Amaro}}, \bibinfo {author} {\bibfnamefont {C.}~\bibnamefont {Maieron}},
  \bibinfo {author} {\bibfnamefont {M.~B.}\ \bibnamefont {Barbaro}}, \bibinfo
  {author} {\bibfnamefont {J.~A.}\ \bibnamefont {Caballero}}, \ and\ \bibinfo
  {author} {\bibfnamefont {T.~W.}\ \bibnamefont {Donnelly}},\ }\href {\doibase
  10.1103/PhysRevC.82.044601} {\bibfield  {journal} {\bibinfo  {journal} {Phys.
  Rev. C}\ }\textbf {\bibinfo {volume} {82}},\ \bibinfo {pages} {044601}
  (\bibinfo {year} {2010})}\BibitemShut {NoStop}%
\bibitem [{\citenamefont {Hern\'andez}\ \emph {et~al.}(2007)\citenamefont
  {Hern\'andez}, \citenamefont {Nieves},\ and\ \citenamefont
  {Valverde}}]{Hernandez:2007qq}%
  \BibitemOpen
  \bibfield  {author} {\bibinfo {author} {\bibfnamefont {E.}~\bibnamefont
  {Hern\'andez}}, \bibinfo {author} {\bibfnamefont {J.}~\bibnamefont {Nieves}},
  \ and\ \bibinfo {author} {\bibfnamefont {M.}~\bibnamefont {Valverde}},\
  }\href {\doibase 10.1103/PhysRevD.76.033005} {\bibfield  {journal} {\bibinfo
  {journal} {Phys. Rev. D}\ }\textbf {\bibinfo {volume} {76}},\ \bibinfo
  {pages} {033005} (\bibinfo {year} {2007})}\BibitemShut {NoStop}%
\bibitem [{\citenamefont {Megias}\ \emph {et~al.}(2015)\citenamefont {Megias},
  \citenamefont {Donnelly}, \citenamefont {Moreno}, \citenamefont {Williamson},
  \citenamefont {Caballero}, \citenamefont {Gonz\'alez-Jim\'enez},
  \citenamefont {De~Pace}, \citenamefont {Barbaro}, \citenamefont {Alberico},
  \citenamefont {Nardi},\ and\ \citenamefont {Amaro}}]{Megias:2014qva}%
  \BibitemOpen
  \bibfield  {author} {\bibinfo {author} {\bibfnamefont {G.~D.}\ \bibnamefont
  {Megias}}, \bibinfo {author} {\bibfnamefont {T.~W.}\ \bibnamefont
  {Donnelly}}, \bibinfo {author} {\bibfnamefont {O.}~\bibnamefont {Moreno}},
  \bibinfo {author} {\bibfnamefont {C.~F.}\ \bibnamefont {Williamson}},
  \bibinfo {author} {\bibfnamefont {J.~A.}\ \bibnamefont {Caballero}}, \bibinfo
  {author} {\bibfnamefont {R.}~\bibnamefont {Gonz\'alez-Jim\'enez}}, \bibinfo
  {author} {\bibfnamefont {A.}~\bibnamefont {De~Pace}}, \bibinfo {author}
  {\bibfnamefont {M.~B.}\ \bibnamefont {Barbaro}}, \bibinfo {author}
  {\bibfnamefont {W.~M.}\ \bibnamefont {Alberico}}, \bibinfo {author}
  {\bibfnamefont {M.}~\bibnamefont {Nardi}}, \ and\ \bibinfo {author}
  {\bibfnamefont {J.~E.}\ \bibnamefont {Amaro}},\ }\href {\doibase
  10.1103/PhysRevD.91.073004} {\bibfield  {journal} {\bibinfo  {journal} {Phys.
  Rev. D}\ }\textbf {\bibinfo {volume} {91}},\ \bibinfo {pages} {073004}
  (\bibinfo {year} {2015})}\BibitemShut {NoStop}%
\bibitem [{\citenamefont {Ivanov}\ \emph {et~al.}(2016)\citenamefont {Ivanov},
  \citenamefont {Megias}, \citenamefont {Gonz\'alez-Jim\'enez}, \citenamefont
  {Moreno}, \citenamefont {Barbaro}, \citenamefont {Caballero},\ and\
  \citenamefont {Donnelly}}]{Ivanov:2015aya}%
  \BibitemOpen
  \bibfield  {author} {\bibinfo {author} {\bibfnamefont {M.~V.}\ \bibnamefont
  {Ivanov}}, \bibinfo {author} {\bibfnamefont {G.~D.}\ \bibnamefont {Megias}},
  \bibinfo {author} {\bibfnamefont {R.}~\bibnamefont {Gonz\'alez-Jim\'enez}},
  \bibinfo {author} {\bibfnamefont {O.}~\bibnamefont {Moreno}}, \bibinfo
  {author} {\bibfnamefont {M.~B.}\ \bibnamefont {Barbaro}}, \bibinfo {author}
  {\bibfnamefont {J.~A.}\ \bibnamefont {Caballero}}, \ and\ \bibinfo {author}
  {\bibfnamefont {T.~W.}\ \bibnamefont {Donnelly}},\ }\href {\doibase
  10.1088/0954-3899/43/4/045101} {\bibfield  {journal} {\bibinfo  {journal}
  {Journal of Physics G: Nuclear and Particle Physics}\ }\textbf {\bibinfo
  {volume} {43}},\ \bibinfo {pages} {045101} (\bibinfo {year}
  {2016})}\BibitemShut {NoStop}%
\bibitem [{\citenamefont {Megias}\ \emph {et~al.}(2017)\citenamefont {Megias},
  \citenamefont {Barbaro}, \citenamefont {Caballero}, \citenamefont {Amaro},
  \citenamefont {Donnelly}, \citenamefont {Ruiz~Simo},\ and\ \citenamefont
  {Van~Orden}}]{Megias:2017cuh}%
  \BibitemOpen
  \bibfield  {author} {\bibinfo {author} {\bibfnamefont {G.~D.}\ \bibnamefont
  {Megias}}, \bibinfo {author} {\bibfnamefont {M.~B.}\ \bibnamefont {Barbaro}},
  \bibinfo {author} {\bibfnamefont {J.~A.}\ \bibnamefont {Caballero}}, \bibinfo
  {author} {\bibfnamefont {J.~E.}\ \bibnamefont {Amaro}}, \bibinfo {author}
  {\bibfnamefont {T.~W.}\ \bibnamefont {Donnelly}}, \bibinfo {author}
  {\bibfnamefont {I.}~\bibnamefont {Ruiz~Simo}}, \ and\ \bibinfo {author}
  {\bibfnamefont {J.~W.}\ \bibnamefont {Van~Orden}},\ }\href@noop {} {\
  (\bibinfo {year} {2017})},\ \Eprint {http://arxiv.org/abs/1711.00771}
  {arXiv:1711.00771 [nucl-th]} \BibitemShut {NoStop}%
\bibitem [{\citenamefont {Simo}\ \emph {et~al.}(2017)\citenamefont {Simo},
  \citenamefont {Amaro}, \citenamefont {Barbaro}, \citenamefont {Caballero},
  \citenamefont {Megias},\ and\ \citenamefont {Donnelly}}]{RUIZSIMO2017193}%
  \BibitemOpen
  \bibfield  {author} {\bibinfo {author} {\bibfnamefont {I.~R.}\ \bibnamefont
  {Simo}}, \bibinfo {author} {\bibfnamefont {J.}~\bibnamefont {Amaro}},
  \bibinfo {author} {\bibfnamefont {M.}~\bibnamefont {Barbaro}}, \bibinfo
  {author} {\bibfnamefont {J.}~\bibnamefont {Caballero}}, \bibinfo {author}
  {\bibfnamefont {G.}~\bibnamefont {Megias}}, \ and\ \bibinfo {author}
  {\bibfnamefont {T.}~\bibnamefont {Donnelly}},\ }\href {\doibase
  https://doi.org/10.1016/j.physletb.2017.04.063} {\bibfield  {journal}
  {\bibinfo  {journal} {Physics Letters B}\ }\textbf {\bibinfo {volume}
  {770}},\ \bibinfo {pages} {193 } (\bibinfo {year} {2017})}\BibitemShut
  {NoStop}%
\bibitem [{\citenamefont {Amaro}\ \emph {et~al.}(1994)\citenamefont {Amaro},
  \citenamefont {C\'{o}},\ and\ \citenamefont {Lallena}}]{AMARO1994365}%
  \BibitemOpen
  \bibfield  {author} {\bibinfo {author} {\bibfnamefont {J.}~\bibnamefont
  {Amaro}}, \bibinfo {author} {\bibfnamefont {G.}~\bibnamefont {C\'{o}}}, \
  and\ \bibinfo {author} {\bibfnamefont {A.}~\bibnamefont {Lallena}},\ }\href
  {\doibase https://doi.org/10.1016/0375-9474(94)90752-8} {\bibfield  {journal}
  {\bibinfo  {journal} {Nuclear Physics A}\ }\textbf {\bibinfo {volume}
  {578}},\ \bibinfo {pages} {365 } (\bibinfo {year} {1994})}\BibitemShut
  {NoStop}%
\bibitem [{\citenamefont {Nieves}\ \emph {et~al.}(2013)\citenamefont {Nieves},
  \citenamefont {Simo},\ and\ \citenamefont {Vacas}}]{NIEVES201390}%
  \BibitemOpen
  \bibfield  {author} {\bibinfo {author} {\bibfnamefont {J.}~\bibnamefont
  {Nieves}}, \bibinfo {author} {\bibfnamefont {I.~R.}\ \bibnamefont {Simo}}, \
  and\ \bibinfo {author} {\bibfnamefont {M.~V.}\ \bibnamefont {Vacas}},\ }\href
  {\doibase https://doi.org/10.1016/j.physletb.2013.03.002} {\bibfield
  {journal} {\bibinfo  {journal} {Physics Letters B}\ }\textbf {\bibinfo
  {volume} {721}},\ \bibinfo {pages} {90 } (\bibinfo {year}
  {2013})}\BibitemShut {NoStop}%
\bibitem [{\citenamefont {Aguilar-Arevalo}\ \emph {et~al.}(2009)\citenamefont
  {Aguilar-Arevalo} \emph {et~al.}}]{PhysRevD.79.072002}%
  \BibitemOpen
  \bibfield  {author} {\bibinfo {author} {\bibfnamefont {A.~A.}\ \bibnamefont
  {Aguilar-Arevalo}} \emph {et~al.} (\bibinfo {collaboration} {MiniBooNE
  Collaboration}),\ }\href {\doibase 10.1103/PhysRevD.79.072002} {\bibfield
  {journal} {\bibinfo  {journal} {Phys. Rev. D}\ }\textbf {\bibinfo {volume}
  {79}},\ \bibinfo {pages} {072002} (\bibinfo {year} {2009})}\BibitemShut
  {NoStop}%
\bibitem [{\citenamefont {Abe}\ \emph {et~al.}(2013)\citenamefont {Abe} \emph
  {et~al.}}]{PhysRevD.87.012001}%
  \BibitemOpen
  \bibfield  {author} {\bibinfo {author} {\bibfnamefont {K.}~\bibnamefont
  {Abe}} \emph {et~al.} (\bibinfo {collaboration} {T2K Collaboration}),\ }\href
  {\doibase 10.1103/PhysRevD.87.012001} {\bibfield  {journal} {\bibinfo
  {journal} {Phys. Rev. D}\ }\textbf {\bibinfo {volume} {87}},\ \bibinfo
  {pages} {012001} (\bibinfo {year} {2013})}\BibitemShut {NoStop}%
\bibitem [{\citenamefont {Aliaga}\ \emph {et~al.}(2016)\citenamefont {Aliaga}
  \emph {et~al.}}]{PhysRevD.94.092005}%
  \BibitemOpen
  \bibfield  {author} {\bibinfo {author} {\bibfnamefont {L.}~\bibnamefont
  {Aliaga}} \emph {et~al.} (\bibinfo {collaboration} {MINER\ensuremath{\nu}A
  Collaboration}),\ }\href {\doibase 10.1103/PhysRevD.94.092005} {\bibfield
  {journal} {\bibinfo  {journal} {Phys. Rev. D}\ }\textbf {\bibinfo {volume}
  {94}},\ \bibinfo {pages} {092005} (\bibinfo {year} {2016})}\BibitemShut
  {NoStop}%
\bibitem [{\citenamefont {Megias}\ \emph {et~al.}(2013)\citenamefont {Megias},
  \citenamefont {Amaro}, \citenamefont {Barbaro}, \citenamefont {Caballero},\
  and\ \citenamefont {Donnelly}}]{Amaro:2013yna}%
  \BibitemOpen
  \bibfield  {author} {\bibinfo {author} {\bibfnamefont {G.~D.}\ \bibnamefont
  {Megias}}, \bibinfo {author} {\bibfnamefont {J.~E.}\ \bibnamefont {Amaro}},
  \bibinfo {author} {\bibfnamefont {M.~B.}\ \bibnamefont {Barbaro}}, \bibinfo
  {author} {\bibfnamefont {J.~A.}\ \bibnamefont {Caballero}}, \ and\ \bibinfo
  {author} {\bibfnamefont {T.~W.}\ \bibnamefont {Donnelly}},\ }\href {\doibase
  10.1016/j.physletb.2013.07.004} {\bibfield  {journal} {\bibinfo  {journal}
  {Phys. Lett.}\ }\textbf {\bibinfo {volume} {B725}},\ \bibinfo {pages} {170}
  (\bibinfo {year} {2013})}\BibitemShut {NoStop}%
\bibitem [{\citenamefont {Nieves}\ \emph
  {et~al.}(2012{\natexlab{b}})\citenamefont {Nieves}, \citenamefont
  {S\'anchez}, \citenamefont {Simo},\ and\ \citenamefont
  {Vacas}}]{PhysRevD.85.113008}%
  \BibitemOpen
  \bibfield  {author} {\bibinfo {author} {\bibfnamefont {J.}~\bibnamefont
  {Nieves}}, \bibinfo {author} {\bibfnamefont {F.}~\bibnamefont {S\'anchez}},
  \bibinfo {author} {\bibfnamefont {I.~R.}\ \bibnamefont {Simo}}, \ and\
  \bibinfo {author} {\bibfnamefont {M.~J.~V.}\ \bibnamefont {Vacas}},\ }\href
  {\doibase 10.1103/PhysRevD.85.113008} {\bibfield  {journal} {\bibinfo
  {journal} {Phys. Rev. D}\ }\textbf {\bibinfo {volume} {85}},\ \bibinfo
  {pages} {113008} (\bibinfo {year} {2012}{\natexlab{b}})}\BibitemShut
  {NoStop}%
\bibitem [{\citenamefont {Martini}\ \emph {et~al.}(2012)\citenamefont
  {Martini}, \citenamefont {Ericson},\ and\ \citenamefont
  {Chanfray}}]{PhysRevD.85.093012}%
  \BibitemOpen
  \bibfield  {author} {\bibinfo {author} {\bibfnamefont {M.}~\bibnamefont
  {Martini}}, \bibinfo {author} {\bibfnamefont {M.}~\bibnamefont {Ericson}}, \
  and\ \bibinfo {author} {\bibfnamefont {G.}~\bibnamefont {Chanfray}},\ }\href
  {\doibase 10.1103/PhysRevD.85.093012} {\bibfield  {journal} {\bibinfo
  {journal} {Phys. Rev. D}\ }\textbf {\bibinfo {volume} {85}},\ \bibinfo
  {pages} {093012} (\bibinfo {year} {2012})}\BibitemShut {NoStop}%
\bibitem [{\citenamefont {Mosel}\ and\ \citenamefont
  {Lalakulich}(2012)}]{Mosel:2012}%
  \BibitemOpen
  \bibfield  {author} {\bibinfo {author} {\bibfnamefont {U.}~\bibnamefont
  {Mosel}}\ and\ \bibinfo {author} {\bibfnamefont {O.}~\bibnamefont
  {Lalakulich}},\ }\href@noop {} {\  (\bibinfo {year} {2012})},\ \Eprint
  {http://arxiv.org/abs/1204.2269} {arXiv:1204.2269 [nucl-th]} \BibitemShut
  {NoStop}%
\bibitem [{\citenamefont {Ankowski}\ \emph {et~al.}(2015)\citenamefont
  {Ankowski}, \citenamefont {Benhar},\ and\ \citenamefont
  {Sakuda}}]{PhysRevD.91.033005}%
  \BibitemOpen
  \bibfield  {author} {\bibinfo {author} {\bibfnamefont {A.~M.}\ \bibnamefont
  {Ankowski}}, \bibinfo {author} {\bibfnamefont {O.}~\bibnamefont {Benhar}}, \
  and\ \bibinfo {author} {\bibfnamefont {M.}~\bibnamefont {Sakuda}},\ }\href
  {\doibase 10.1103/PhysRevD.91.033005} {\bibfield  {journal} {\bibinfo
  {journal} {Phys. Rev. D}\ }\textbf {\bibinfo {volume} {91}},\ \bibinfo
  {pages} {033005} (\bibinfo {year} {2015})}\BibitemShut {NoStop}%
\bibitem [{\citenamefont {Lalakulich}\ and\ \citenamefont
  {Mosel}(2013)}]{PhysRevC.87.014602}%
  \BibitemOpen
  \bibfield  {author} {\bibinfo {author} {\bibfnamefont {O.}~\bibnamefont
  {Lalakulich}}\ and\ \bibinfo {author} {\bibfnamefont {U.}~\bibnamefont
  {Mosel}},\ }\href {\doibase 10.1103/PhysRevC.87.014602} {\bibfield  {journal}
  {\bibinfo  {journal} {Phys. Rev. C}\ }\textbf {\bibinfo {volume} {87}},\
  \bibinfo {pages} {014602} (\bibinfo {year} {2013})}\BibitemShut {NoStop}%
\bibitem [{\citenamefont {Betancourt}\ \emph {et~al.}(2017)\citenamefont
  {Betancourt} \emph {et~al.}}]{PhysRevLett.119.082001}%
  \BibitemOpen
  \bibfield  {author} {\bibinfo {author} {\bibfnamefont {M.}~\bibnamefont
  {Betancourt}} \emph {et~al.} (\bibinfo {collaboration}
  {MINER\ensuremath{\nu}A Collaboration}),\ }\href {\doibase
  10.1103/PhysRevLett.119.082001} {\bibfield  {journal} {\bibinfo  {journal}
  {Phys. Rev. Lett.}\ }\textbf {\bibinfo {volume} {119}},\ \bibinfo {pages}
  {082001} (\bibinfo {year} {2017})}\BibitemShut {NoStop}%
\bibitem [{\citenamefont {Benhar}\ \emph {et~al.}(1994)\citenamefont {Benhar},
  \citenamefont {Fabrocini}, \citenamefont {Fantoni},\ and\ \citenamefont
  {Sick}}]{Benhar:1994hw}%
  \BibitemOpen
  \bibfield  {author} {\bibinfo {author} {\bibfnamefont {O.}~\bibnamefont
  {Benhar}}, \bibinfo {author} {\bibfnamefont {A.}~\bibnamefont {Fabrocini}},
  \bibinfo {author} {\bibfnamefont {S.}~\bibnamefont {Fantoni}}, \ and\
  \bibinfo {author} {\bibfnamefont {I.}~\bibnamefont {Sick}},\ }\href {\doibase
  10.1016/0375-9474(94)90920-2} {\bibfield  {journal} {\bibinfo  {journal}
  {Nucl. Phys.}\ }\textbf {\bibinfo {volume} {A579}},\ \bibinfo {pages} {493}
  (\bibinfo {year} {1994})}\BibitemShut {NoStop}%
%%CITATION = NUPHA,A579,493;%%
\end{thebibliography}%

\end{document}